\documentclass[aps,pra,reprint,showkeys,onecolumn,floatfix,showpacs]{revtex4-2}
\usepackage{multirow}
\usepackage{graphicx}
\usepackage{enumerate}
\usepackage{amssymb}
\usepackage{fdsymbol}
\usepackage{mathtools}
\usepackage{wasysym}

\begin{document}

\newtheorem{theorem}{Theorem}[section]
\newtheorem{definition}[theorem]{Definition}
\def\crta{\vrule height1.41ex depth-1.27ex width0.34em}
\def\dj{d\kern-0.36em\crta}
\def\Crta{\vrule height1ex depth-0.86ex width0.4em}
\def\Dj{D\kern-0.73em\Crta\kern0.33em}
%\dimen0=\hsize \dimen1=\hsize \advance\dimen1 by 40pt
\title{Non-Kochen-Specker Contextuality}
\author{Mladen Pavi{\v c}i{\'c}}
\email[]{mpavicic@irb.hr}
\affiliation{Center of Excellence for Advanced Materials 
and Sensing Devices (CEMS), Photonics and Quantum Optics Unit, 
Ru{\dj}er Bo\v skovi\'c Institute 
and Institute of Physics, Zagreb, Croatia.}
%\date{\today}

\begin{abstract}{Quantum contextuality supports quantum computation
  and communication. One of its main vehicles is hypergraphs.
  The most elaborated are the Kochen-Specker ones, but there is also
  another class of contextual sets that are not of this kind.
  Their representation has been mostly operator-based and limited
  to special constructs in three- to six-dim spaces, a notable
  example of which is the Yu-Oh set. Previously, we showed
  that hypergraphs underlie all of them, and in this paper, we
  give general methods---whose complexity does not scale up with the
  dimension---for generating such non-Kochen--Specker hypergraphs
  in any dimension and give examples in up to 16-dim spaces.
  Our automated generation is probabilistic and random, but the
  statistics of accumulated data enable one to filter out sets
  with the required size and structure.}
\end{abstract}


\maketitle

\section{Introduction}

Quantum contextuality, which precludes assignments of predetermined
values to dense sets of states, has found applications in quantum
communication \cite{cabello-dambrosio-11,nagata-05,saha-hor-19},
quantum computation \cite{magic-14,bartlett-nature-14}, quantum
nonlocality \cite{kurz-cabello-14},  quantum steering
\cite{tavakoli-20}, and lattice theory
\cite{bdm-ndm-mp-fresl-jmp-10,mp-7oa}. Small contextual set
experiments were carried out with photons
\cite{simon-zeil00,michler-zeil-00,amselem-cabello-09,liu-09,d-ambrosio-cabello-13,ks-exp,ks-exp-03,lapkiewicz-11,zu-wang-duan-12,canas-cabello-8d-14,canas-cabello-14,zhan-sanders-16},
classical light \cite{li-zeng-17,li-zeng-zhang-19,frustaglia-cabello-16,zhang-zhang-19},
neutrons \cite{h-rauch06,cabello-fillip-rauch-08,b-rauch-09},
trapped ions \cite{k-cabello-blatt-09},
solid state molecular nuclear spins \cite{moussa-09},
and superconducting quantum systems \cite{jerger-16}.

There are three classes of contextual sets elaborated on in the
literature which are not of the more common kind of
Kochen--Specker (KS) sets
\cite{koch-speck,pavicic-pra-22,pavicic-quantum-23}
and for which we provide a hypergraph generalization in
this paper. 

The first class consists of the operator-based
state-independent contextual (SIC) sets put forward by Klyachko
et al.~\cite{klyachko-08}, Yu and Oh \cite{yu-oh-12},
Bengtsson, Blanchfield, and Cabello \cite{beng-blan-cab-pla12},
Xu, Chen, and Su \cite{xu-chen-su-pla15}, Ramanathan and
Horodecki \cite{ram-hor-14}, and Cabello, Kleinmann, and Budroni
\cite{cabell-klein-budr-prl-14}, which are not Kochen--Specker
sets. 

The second class consists of hypergraphs built by multiples of
mutually orthogonal vectors where at least one of the
multiples contains less than $n$ vectors, where $n$ is the
dimension of space in which a hypergraph resides
\cite{magic-14,pavicic-entropy-19,pavicic-quantum-23}. 

The third class consists of the so-called true-implies-false
and true-implies-true sets \cite{cabello-svozil-18,svozil-21}.

All sets from these three classes as well as their hypergraph
generalization that we elaborate on are contextual, 
and therefore, we call them non-KS contextual sets. 

We provide a general method for arbitrarily generating many non-KS
hypergraphs in spaces of up to 16-dim. In order to achieve these
goals, we make use of non-binary non-KS
McKay--Megill--Pavi{\v c}i{\'c} hypergraphs (MMPHs) and their
language. By means of our algorithms and programs, we arbitrarily
obtain many MMPHs, which can be used for various
applications, e.g., to generate new entropic tests of
contextuality or new operator-based contextual sets. 

The paper is organized as follows.

In Section \ref{subsec:form}, we present the hypergraph language
and formalism and define non-binary MMPHs (NBMMPH) and binary
MMPHs (BMMPH). We explain how vertices and hyperedges in
an MMPH and in $n$-dim space correspond to vectors and their
orthogonalities, i.e., $m$-tuples ($2\le m\le n$) of mutually
orthogonal vectors, respectively.

In Section \ref{subsec:nonks}, we present three methods of generating
non-KS MMPHs.

In Section \ref{subsec:classes3-5}, we give examples of the
aforementioned non-KS sets.

In Sections \ref{subsec:classes3-5}--\ref{subsec:classes6-8},
we generate four- to eight-dim critical non-KS NBMMPHs from master sets,
themselves generated from simple vector components.

In Sections \ref{subsec:classes9-11}--\ref{subsec:classes12-16},
we obtain nine- to sixteen-dim critical non-KS NBMMPHs via the dimensional
upscaling method, which does not scale up with dimension.

In Section \ref{sec:disc}, we discuss and review the steps and details
of our methods. 

In Section \ref{sec:met}, we give the technical methods used in the paper.

In Section \ref{sec:concl}, we summarize the results achieved in the
paper.

\section{\label{sec:results}Results}

We consider a set of quantum states represented by vectors in
$n$-dim Hilbert space ${\cal H}^n$ grouped into $m$-tuples ($m\le n$)
of mutually orthogonal vectors with $m<n$ holding for at least
one $m$. We describe such a set by means of MMPHs. In it, vectors
themselves are represented by vertices and mutually orthogonal
$m$-tuples of them by hyperedges. However, an MMPH itself has a
definition that is independent of a possible representation
of vertices by means of vectors. For instance, there are MMPHs
without coordinatization, i.e., MMPHs for whose vertices, vectors
do not exist. When coordinatization exists, that does not mean
that $n-m$ vertices in considered hyperedges do not or cannot
exist, but only that we do not take the remaining $n-m$
vertices/vectors into account while elaborating on properties
of vertices and hyperedges.

\subsection{\label{subsec:form}Formalism}

Let us define the MMPH formalism/language \cite{pavicic-quantum-23}.

\begin{definition}\label{def:mmp} An {\rm MMPH} is an $n$-dim%MDPI: Please check if it should keep the form of mormal or should be changed into italic, please check check it for the whole paper.
%%Author: no italics; normal, upright, roman font throughout the paper, for the whole paper
  
hypergraph $k$-$l$ with $k$ vertices and $l$ hyperedges in which
\begin{enumerate}
\item Every vertex belongs to at least one hyperedge;
\item Every hyperedge contains at least two and at most $n$ vertices;
\item No hyperedge shares only one vertex with
  another hyperedge;
\item Hyperedges may intersect each other in at most $n-2$
  vertices
\item Graphically, vertices are represented as dots, and
  hyperedges are (curved) lines passing through them.
\end{enumerate}
\end{definition}

\begin{definition}\label{df:nbmmph}
  An $n$-dim {\rm non-binary MMPH (NBMMPH)}, $n\ge 3$,
  {\rm \cite{pm-entropy18}} is an {\rm MMPH} for which each hyperedge
  contains $m$ vertices, $2\le m\le n$, and to which it is
  impossible to assign $1$ s and $0$ s in such a way that
  \begin{enumerate}
\item No two vertices within any of its edges are both assigned
  a value of $1$;
\item In any of its edges, not all of the vertices are
  assigned a value of $0$.
\end{enumerate}
\end{definition}

\begin{definition}\label{df:ks}
  An {\rm NBMMPH} in which $m=n$ holds for all
  hyperedges is a {\rm KS MMPH}.
\end{definition}

For $m\>$=$\>n$, an {\rm {NBMMPH}} reduces to a {\rm {KS}} contextuality
set, i.e., to a set satisfying the Kochen--Specker theorem
 \cite{koch-speck,budroni-cabello-rmp-22,zimba-penrose,pavicic-quantum-23}.

\begin{definition}\label{df:nonks}
  An {\rm NBMMPH} in which $m<n$ holds for at least one
  hyperedge is a {\rm non-KS MMPH}.
\end{definition}

In this paper, we consider only those non-KS MMPHs for
which $m=n$ for at least one hyperedge.

\begin{definition}\label{df:bmmph}
  An $n$-dim {\rm binary MMPH (BMMPH)}, $n\ge 3$,
  is an {\rm MMPH} for which each hyperedge contains $m$ vertices,
  $2\le m\le n$, and to which it is possible to assign 1 s and
  0 s in such a way that
  \begin{enumerate}
\item No two vertices within any of its edges are both assigned
  a value of 1;
\item In any of its edges, not all of the vertices are
  assigned a value of 0.
\end{enumerate}
\end{definition}

\begin{definition}\label{df:filled}
  A {\rm non-KS NBMMPH} to which vertices are added so as to
  make the number of vertices equal to $n$ in every hyperedge is
  called a {\rm filled MMPH}.
\end{definition}

Filled MMPHs are mostly BMMPHs.

\begin{definition}\label{df:critical} A {\em critical} {\rm NBMMPH}
  is an {\rm NBMMPH} that is minimal in the sense that
  removing any of its hyperedges turns it into a {\rm BMMPH}.
\end{definition}

\begin{definition}\label{df:multi} Vertex multiplicity is
  the number of hyperedge vertexes ``$i$'' belongs to; we denote
  it by $m(i)$.
\end{definition}

\begin{definition}\label{df:class} A {\em master} is a non-critical
  {\rm MMPH} that contains smaller critical and non-critical
  sub-{\rm MMPH}s. A collection of sub-{\rm MMPH}s of an {\rm MMPH}
  master forms its {\em class}. 
\end{definition}

  A {\em parity proof} of
  the contextuality of a $k$-$l$ NBMMPH with odd $l$ and where each
  vertex shares an even number of edges stems from its
  inherent contradiction: because each vertex shares
  an even number of  hyperedges, there should be an even
  number of hyperedges with 1s. At the same time, each edge
  can contain only one 1 by definition, and since there are
  an odd number of hyperedges in the MMPH, there should also
  be an odd number of edges with 1s 

\begin{definition}\label{df:coord}
  A {\em coordinatization} of a non-{\rm KS} {\rm NBMMPH} is a set
  of vectors assigned to its vertices that is a subset of $n$-dim
  vectors in ${\cal H}^n$, $n\ge 3$, assigned to vertices
  of its filled {\rm MMPH} or its smallest master (they need not
  coincide) or any of its masters.
\end{definition}

In other words, a ``coordinatization'' of each hyperedge of a
filled MMPH or a smallest master MMPH is represented by an
$n$-tuple of orthogonal vectors, while a ``coordinatization'' of
each hyperedge of the original non-KS NBMMPH is represented by a
vector $m$-tuple ($m\le n$), which is a subset of that $n$-tuple.
This means that the former MMPH inherits its coordinatization
from the coordinatization of its master or its filled set
(they may, but usually do not, coincide) or any its masters.
In our present approach, a coordinatization is automatically
assigned to each hypergraph by the very procedure of its
generation from master MMPHs, as we show below. 

An MMPH is encoded with the help of printable ASCII characters,
with the exception of ``space'', ``0'', ``+'', ``,'' and ``.'',
organized in single strings; its hyperedges are separated by
commas, and each string ends with a period.  When all ASCII
characters are exhausted, one reuses them prefixed by ``+'',
and then again by ``++'', and so forth. An MMPH with $k$
vertices and $l$ edges is denoted as a $k$-$l$ MMPH. ASCII string
representation is used for computer processing. MMPH strings are
handled by means of algorithms embedded in the programs
SHORTD, MMPSTRIP, MMPSUBGRAPH, VECFIND, STATES01, and others
 \cite{bdm-ndm-mp-1,pmmm05a-corr,pmm-2-10,bdm-ndm-mp-fresl-jmp-10,mfwap-s-11,mp-nm-pka-mw-11}.

\subsection{\label{subsec:nonks}Generation of
  Non-KS MMPHs}

To generate non-KS NBMMPHs, we make use of the following methods.

\begin{itemize} 
\item{\bf M1} 
consists of dropping vertices contained in single
hyperedges (multiplicity $m=1$) \cite{pavicic-quantum-23}
of either NBMMPHs or BMMPHs and a possible subsequent stripping
of their hyperedges. The obtained smaller MMPHs are often
non-KS, although never KS.  
\item{\bf M2} consists of a random addition of hyperedges to
MMPHs so as to obtain bigger ones, which then serve us to generate
smaller non-KS NBMMPHs by stripping hyperedges randomly again;
\item{\bf M3} consists of the random deletion of vertices in either
NBMMPHs or a BMMPHs until a non-KS NBMMPH is reached.
\end{itemize}

We combine all three methods to obtain an arbitrary
number of non-KS NBMMPHs in an arbitrary dimension.
The methods rely on the property of MMPHs where, by stripping
an MMPH or NBMMPH (critical or not) or BMMPH of its hyperedges,
we can arrive at smaller non-KS NBMMPHs in contrast to a
critical KS NBMMPH whose stripping of hyperedges can never
yield another (smaller) NBMMPH. 

\subsection{\label{subsec:classes3-5}Dimensions Three to Five and
  the Three Classes of Non-KS Contextual Sets from the
  Literature}

\begin{figure}[ht]
\begin{center}
  \includegraphics[width=0.99\textwidth]{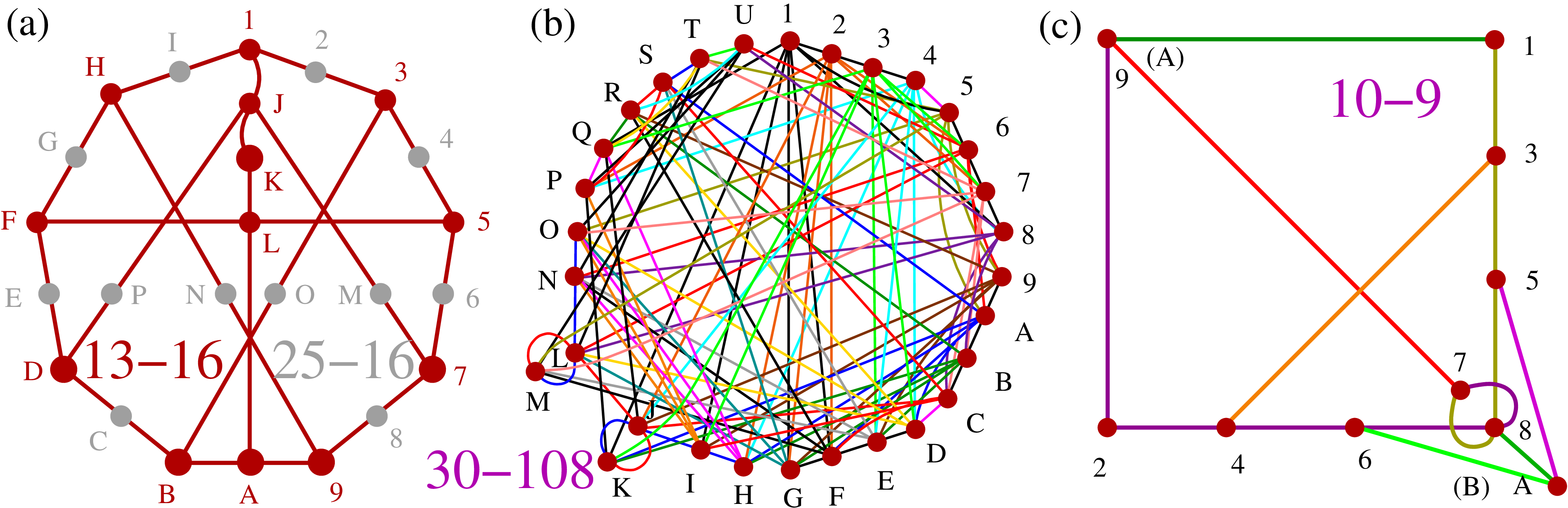}
\end{center}
\caption{(\textbf{a}) Yu-Oh's three-dim non-KS 13-16 non-KS NBMMPH
  (\cite[Figure 2]{yu-oh-12}); 
  gray vertices that enlarge 13-16 to 25-16 are necessary for
  coordinatization and implementation; 
  (\textbf{b}) Howard, Wallman, Veitech, and Emerson's four-dim 30-108
  non-KS NBMMPH  (\cite[Figure 2]{magic-14}); (\textbf{c}) Cabello, Portillo,
  Sol\'is, and Svozil's five-dim 10-9 non-KS NBMMPH
  (\cite[Figure 5a]{cabello-svozil-18}); the original symbols are presented in
  brackets (A,B).}
\label{fig:3-5}
\end{figure}

In Figure \ref{fig:3-5}, we give examples from each of the
three classes of non-KS sets referred to in the
Introduction. Here, we remind the reader that $k$-$l$ MMPHs
refer to hypergraphs with $k$ vertices and $l$ hyperedges
(Definition \ref{def:mmp}), while the corresponding graphs have more
than $l$ edges. For example, in Figure \ref{fig:3-5}a, the hypergraph
hyperedge {\tt ALK} corresponds to a graph clique with three edges:
{\tt AL}, {\tt LK}, and {\tt KA}. 

Yu-Oh's three-dim non-KS NBMMPH, shown in Figure \ref{fig:3-5}a, is
presumably the earliest of the kind. It is operator-based, but
the operators are defined via states/vectors/vertices of 13-16
MMPH, as reviewed in \cite{pavicic-entropy-19}. Since orthogonal
vectors in a three-dim space form triples, full representation
requires 25-16, as indicated by the gray vertices in the figure,
which can be obtained from Peres' 33-40 \cite{pavicic-entropy-19}
by stripping hyperedges and the 13-16 from it by removing $m=1$
vertices, i.e., via {\bf M1}. The 13-16 MMPH is not critical, and it
contains four critical sub-MMPHs, the smallest of which is 10-9
 \cite{pavicic-entropy-19}.

Howard, Wallman, Veitech, and Emerson's four-dim 30-108 non-KS NBMMPH,
shown in Figure \ref{fig:3-5}b, which was obtained from the set of stabilizer
states was used to prove that the underlying contextuality is
essential for quantum computation. We discuss its filled 232-108
MMPH and its critical 24-71 MMPH in \cite{pavicic-quantum-23}.

Cabello, Portillo, Sol\'is, and Svozil's five-dim 10-9 non-KS NBMMPH,
shown in Figure \ref{fig:3-5}b, is one of the minimal five-dim
true-implies-false sets (TIFS) (\cite[Figure 5a]{cabello-svozil-18}).
It is not critical, and the only critical part it contains is a 10-7, but
it is not a TIFS any more. The coordinatization of the filled 10-9
(31-9, which includes the coordinatization of 10-9 itself) can be
built from the $\{0,\pm1,2\}$ components and is given in
Appendix \ref{app:1a}.

Our methods can generate NBMMPHs that are critical as well as
those that are not. Therefore, although none of the
aforementioned examples are critical, we focus on critical
ones, because they offer the simplest implementation and 
presentation. The rationale for adopting such an approach
is that only minimal contextual sets, i.e., critical NBMMPHs,
are relevant for experimental implementations, since their
supersets just contain additional orthogonalities that do not
change the contextuality property of their smallest critical set.
Hence, while designing MMPHs for particular implementations, we
should attempt to find the ones that are critical and are provided
via automated generations of MMPHs.    

\begin{figure}[ht]
\begin{center}
  \includegraphics[width=0.99\textwidth]{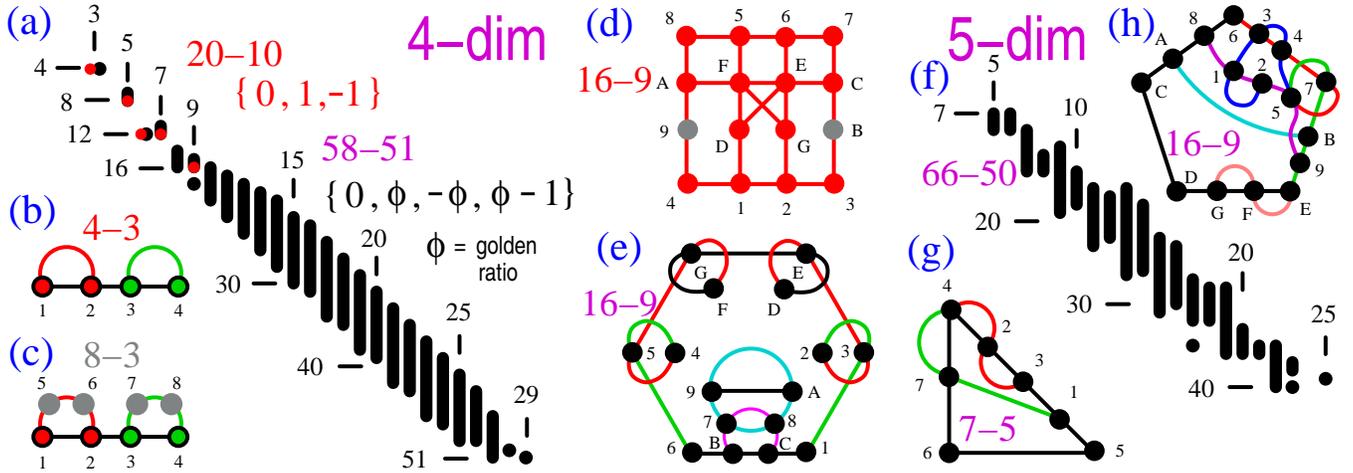}
\end{center}
\caption{(\textbf{a}) Distributions of critical four-dim non-KS NBMMPHs
  obtained from submaster 20-10, which was obtained from (Peres')
  24-24 supermaster (generated by vector components $\{0,\pm1\}$)
  by {\bf M1} (dots in red) and from submaster
  58-51, itself obtained from the 60-72 supermaster (generated by
  vector components $\{0,\pm\phi,\phi-1\}$, where $\phi$ is the
  golden ratio: $\frac{1+\sqrt{5}}{2}$) by {\bf M1} (in
  black); the abscissa is $l$ (number of hyperedges); and the ordinate is $k$
  (number of vertices). The dots represent $(k,l)$. Consecutive dots
  (same $l$) are shown as strips;
 (\textbf{b}) the smallest non-KS in the distributions: 4-3;
  (\textbf{c}) BMMPH 8-3---filled with 4-3---which one needs for obtaining the
  coordinatization and implementation of 4-3; (\textbf{d}) the 16-9
  critical obtained from the 20-10 master; (\textbf{e}) the 16-9
  critical obtained from the 58-51 master; (\textbf{f}) distributions of
  critical five-dim non-KS NBMMPHs obtained from submaster 66-50
  which was obtained from the 105-136 supermaster (generated by
  vector components $\{0,\pm1\}$); (\textbf{g}) the smallest critical;
  (\textbf{h}) a 16-9 critical for the sake of comparison with four-dim
  16-9s; strings and coordinatizations are given in Appendix \ref{app:1a}.}
\label{fig:3d-5d}
\end{figure}

In \cite{pavicic-entropy-19}, we give ample distributions of
three-dim non-KS NBMMPHs obtained via {\bf M1} and {\bf M2}.
Therefore, below, we give distributions and samples of just
four- and five-dim critical non-KS NBMMPHs presented in
Figure \ref{fig:3d-5d}a,f. Here, we only point out
that the KS ``bug,'' the 8-7 non-KS NBMMPH shown in
(\cite[Figure 3a]{pavicic-entropy-19}), is the smallest three-dim 
non-KS NBMMPH that satisfies our requirement that at least one
of the hyperedges must contain $n$ vertices ($n$ being the dimension
of the considered MMPH), none of which has the multiplicity
$m=1$. Its string, the string of its filled MMPH, and their 
coordinatizations are given in Appendix \ref{app:1-3}, as are the
strings and coordinatizations of any other MMPH considered in
the paper given in Appendix \ref{app:1}. 

To obtain non-KS NBMMPHs via {\bf M1}, we first generate the
supermasters from the vector components. In the four-dim space, we
obtain the 24-24 supermaster from the $\{0,\pm1\}$ components
and the 60-72 supermaster from the $\{0,\pm\phi,\phi-1\}$
components, where $\phi=\frac{1+\sqrt{5}}{2}$ (the golden
ratio). Their strings and coordinatizations are given in
Appendix \ref{app:1a}. Then, we randomly strip hyperedges from them,
e.g., 14 from 24-24 and 21 from the 60-72 supermaster, so as to
obtain the 20-10 and 58-51 masters, respectively. From the latter
masters, we remove $m=1$ vertices, and from any of them, we generate
the classes of critical MMPHs by stripping them further until we
obtain critical MMPHs that form the 20-10 and 58-51 non-KS classes.
In the five-dim space, we obtain the 105-136 supermaster from the
$\{0,\pm1\}$ components. Its string and coordinatization are given
in Appendix \ref{app:1b}. Further, we randomly strip 86 hyperedges to
obtain a 66-50 master and eventually obtain its class of critical non-KS
NBMMPHs. 

We generate $n$-dim critical non-KS MMPHs under the requirement that
at least one of their hyperedges must contain $n$ vertices, of
which none have a multiplicity of 1 ($m=1$). (All examples from
Figure \ref{fig:3-5} satisfy these conditions.) For instance,
the smallest critical obtained in the four-dim distribution, shown in
Figure \ref{fig:3d-5d}a, is the 4-3 shown in Figure \ref{fig:3d-5d}b,
whose hyperedge {\tt 1234} is of such a kind. Its filled MMPH shown
in Figure \ref{fig:3d-5d}c provides a coordinatization
necessary for the implementation of the 4-3. The 16-9 critical of the
20-10 master shown in Figure \ref{fig:3d-5d}(d)
contains two $m=1$ vertices ({\tt 9,B}), because $m=1$ vertices
were stripped only once (from the master) when we started the
generation of the 20-10 class. We can remove one or both of these
vertices and still have a critical non-KS MMPH (15-9 or 14-9,
respectively) if we want to for some reason. The 16-9 critical
shown in Figure \ref{fig:3d-5d}e has a parity proof, since in it,
each vertex shares exactly two hyperedges, while there is an odd
number of them (9). Strings and coordinatizations are given in
Appendix \ref{app:1b}.

\subsection{\label{subsec:classes6-8}Dimensions Six to Eight}

Cabello, Portillo, Sol\'is, and Svozil also give a number of minimal
six-dim TIFS non-KS NBMMPHs in (\cite[Figure 7]{cabello-svozil-18})
along the same line as for their five-dim one shown in
Figure \ref{fig:3-5}c. To our knowledge, there are no explicit
examples of non-KS NBMMPH in dimensions seven and eight in the
literature. Therefore, we straightforwardly move to the generation
of six- to eight-dim non-KS NBMMPHs.

An NBMMPH in the six-dim Hilbert space corresponds to a qubit entangled
with a qutrit (${\cal H}^6={\cal H}^2\otimes{\cal H}^3$) or  a
$\frac{5}{2}$-spin system. So far, to obtain KS NBMMPH masters,
the following vector components have been used: $\{0,\pm\omega\}$,
\cite{pavicic-pra-17,pwma-19,pm-entropy18} ($\omega$ is a cube
root of 1, $\omega=e^{2\pi i/3}=(i\sqrt{3}-1)/2$),
$\{0,\pm\omega,\omega^2\}$ \cite{lisonek-14,pwma-19} and
$\{0,\pm1\}$ \cite{pavicic-pra-17}. Since the first set of
components yields a master with only three MMPHs, we
make use of the other two to generate six-dim non-KS NBMMPHs. 

\begin{figure}[ht]
\begin{center}
  \includegraphics[width=\textwidth]{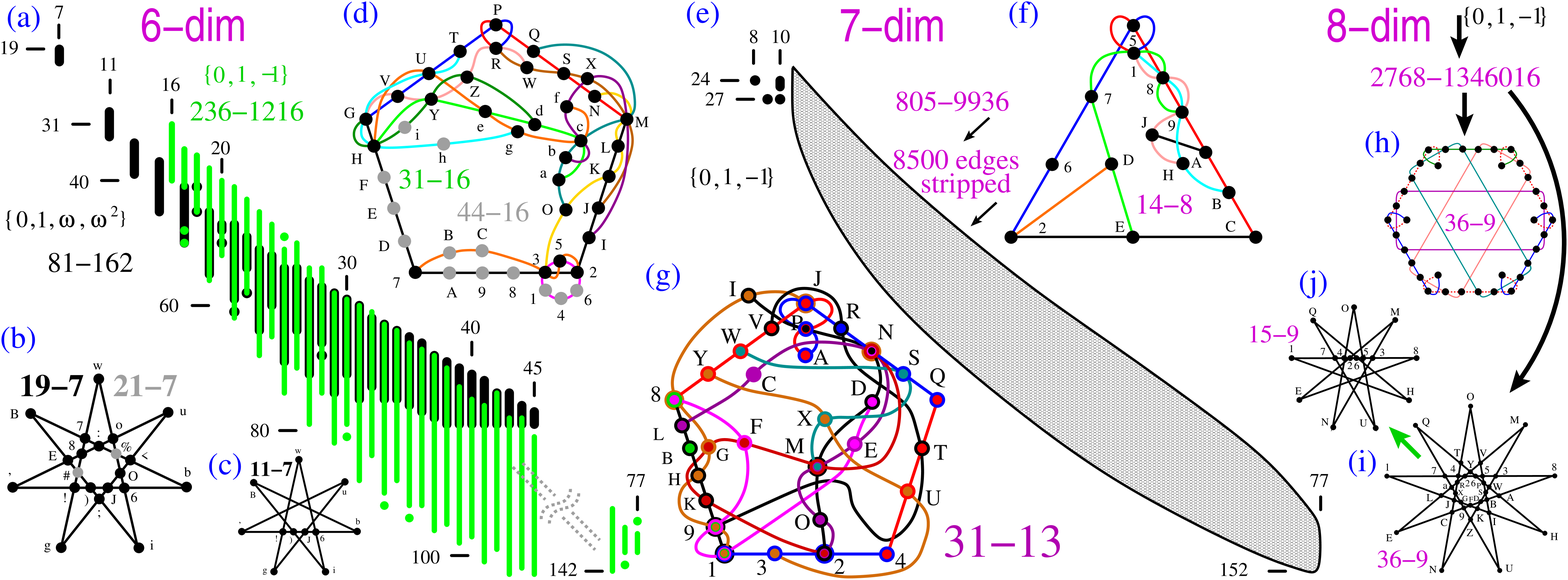}
\end{center}
\caption{(\textbf{a}) Distributions of six-dim critical non-KS
  NBMMPHs obtained from two different submasters---see text;
  (\textbf{b}) the smallest critical non-KS NBMMPH obtained
  from the former class by {\bf M3}, which has a parity proof;
  (\textbf{c}) an even smaller critical non-KS NBMMPH obtained
  from it by hand, which has a parity proof; (\textbf{d}) the
  smallest critical non-KS NBMMPH obtained from the latter
  class by {\bf M1}; (\textbf{e}) distributions of seven-dim
  critical non-KS NBMMPHs---see text; (\textbf{f}) 14-8 non-KS
  NBMMPH, one of the smallest non-KS NBMMPHs obtained via {\bf M3}
  from the smallest KS NBMMPH, 34-14; (\textbf{g}) 31-13 also
  obtained from the 34-14 (no $m=1$ vertices essential for
  criticality); (\textbf{h},\textbf{i}) two 8-dim KS MMPHs with
  the smallest number of hyperedges (\textbf{9}); (\textbf{i});
  serves us in generating the 15-9 non-KS NBMMPH in (\textbf{j});
  (\textbf{h}--\textbf{j}) MMPHs have parity proofs; strings and
  coordinatizations are given in Appendices
  \ref{app:1c}--\ref{app:1e}.}
\label{fig:6d-8d}
\end{figure}

The $\{0,1,\omega,\omega^2\}$ set generates two unconnected
masters: 591-1123 and 81-162 \cite{pwma-19}. To obtain non-KS
NBMMPHs, we apply {\bf M1} to the 81-162 class. Their distribution
is shown in Figure \ref{fig:6d-8d}a in black. The $\{0,\pm 1\}$ set
generates a 236-1216 master. Its non-KS NBMMPHs are also obtained
via {\bf M1} and are shown in Figure \ref{fig:6d-8d}a in green. 

In the seven-dim space, masters obtained from simple vector components,
such as $\{0,\pm 1\}$, are too big to be used for the exhaustive
generation of a complete non-KS NBMMPH class.
Instead, as in the previous six-dim case, we strip a significant
portion of hyperedges from a master obtained from $\{0,\pm 1\}$
components and make use of the remaining MMPHs to obtain a
non-KS class, as shown in Figure \ref{fig:6d-8d}e; $\{0,\pm 1\}$
yields the 805-9936 master, and stripping of 8500 hyperedges leaves
us with NBMMPHs with 436 hyperedge NBMMPHs, which generates a 
436-hyperedge class. Since this class is still big, we have to
repeat {\bf M1} several times to obtain small non-KS critical
NBMMPHs. As a result, hyperedges of all small NBMMPHs may contain
some $m=1$ vertices essential for criticality, as shown in
Figure \ref{fig:6d-8d}f (the removal of vertex {\tt 6} would terminate
the criticality of the MMPH). In dimensions greater than nine, such
vertices do not appear, although even here we can avoid their
generation by applying {\bf M3} to KS NBMMPHs, as shown in
Figure \ref{fig:6d-8d}g.

The eight-dim MMPH master is big (2768-1346016), but the stripping
technique can still provide us with non-KS NBMMPHs via {\bf M1}.
However, the MMPHs with $m=1$ vertices are also big, and
obtaining small criticals with up to 40 hyperedges would
require roughly one week on a supercomputer with
200 2.5 GHz CPUs working in parallel.
We may be able to work around this problem by exploiting previously
generated small KS criticals \cite{pavicic-pra-17} so as to
use them as masters for non-KS MMPHs while applying {\bf M3},
as shown in Figure \ref{fig:6d-8d}h--j (cf. the six-dim star
in Figure \ref{fig:6d-8d}b). Notice the
graphical similarity of the four-dim 18-9 (\cite[Figure 3a]{pmmm05a})
and eight-dim 36-9 (shown in Figure \ref{fig:6d-8d}h) for each vertex
from the 18-9 vs. a pair of vertices in the 36-9. 
Since the
distribution of eight-dim KS MMPHs in Ref. \cite{pavicic-pra-17}
is abundant, we can arbitrarily generate many non-KS NBMMPHs
in this manner via {\bf M3}. 

\subsection{\label{subsec:classes9-11}Dimensions Nine to Eleven}

\begin{figure}[ht]
\begin{center}
  \includegraphics[width=\textwidth]{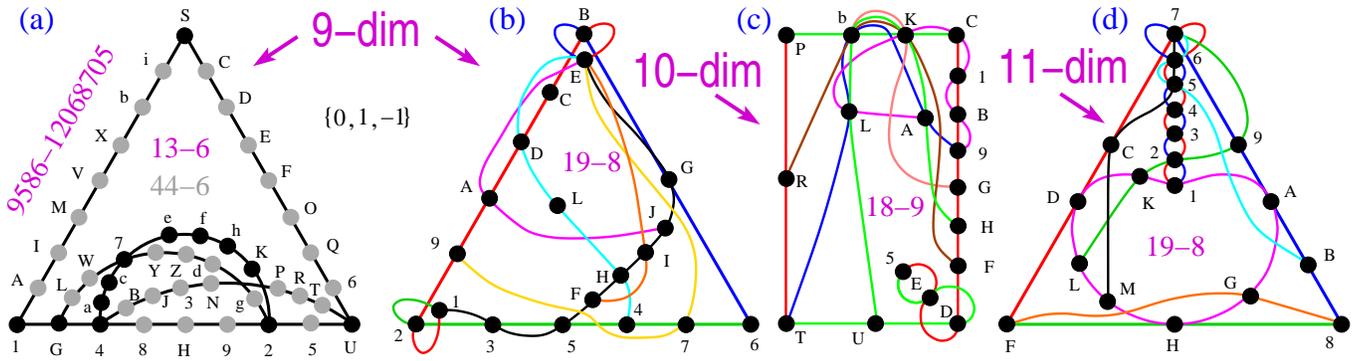}
\end{center}
\caption{(\textbf{a}) The 44-6 BMMPH and its critical subgraph 13-6
  non-KS NBMMPH directly obtained from the supermaster via {\bf M1};
  (\textbf{b}) the critical nine-dim 19-8 obtained via {\bf M3} from
  the master 47-16; (\textbf{c}) the critical ten-dim 18-9 non-KS
  NBMMPH obtained via {\bf M3} from the 50-15 master; (\textbf{d})
  the critical eleven-dim 19-8 non-KS NBMMPH obtained via {\bf M3}
  from the 50-14 master. Strings and coordinatizations are given in
  Appendices \ref{app:1f}--\ref{app:1h}.}
\label{fig:9d-11d}
\end{figure}

The nine-dim NBMMPH master obtained from $\{0,\pm1\}$ has 9586
vertices and 12,068,705 hyperedges and that is too big
for the direct generation of critical MMPHs (via stripping and
filtering), especially for higher dimensions. However, billions
of BMMPHs can be generated from the master, and as we have already
stressed, stripping them of $m=1$ often provides us with NBMMPHs.
This renders {\bf M1} applicable. Thus, after the random stripping of
12,068,200 hyperedges, we obtained submasters with 505 hyperedges.
By requiring that at least one of the hyperedges contains $n$ vertices
and that some of them can have the multiplicity $m=1$, our program
STATES01 yields a series of critical NBMMPHs, the smallest of which
is 13-6, as shown in Figure \ref{fig:9d-11d}a. The hyperedge
{\tt 4ac7efhK2} contains nine vertices. (Notice also that
the 13-6 NBMMPH remains a critical non-KS NBMMPH with any,
some, or all of {\tt a,c,e,f,h,K} removed.) The filled 13-16,
i.e., 44-6, also shown in Figure \ref{fig:9d-11d}a, obtains
the coordinatization directly from the supermaster, since the
programs preserve the names of the vertices in the process
of stripping and yielding sub-MMPHs. Obtaining a
coordinatization via VECFIND takes too many CPU hours. The latter feature also makes {\bf M2} inapplicable.

If we wanted to keep our $n$-vertex requirement in full (``no
$m=1$ vertices''), in order to obtain critical non-KS NBMMPHs,
we would need to employ {\bf M3}, so as to apply it on KS NBMMPHs
obtained via dimensional upscaling \cite{waeg-aravind-pra-17,pw-23},
as follows. We removed several vertices from the smallest critical
47-16 obtained in \cite{pw-23} until it was not critical any
more. Then, STATES01 yielded the 19-8 critical shown in
Figure \ref{fig:9d-11d}b. (The removal of vertex {\tt L} would
terminate the criticality of the MMPH as with the seven-dim one
shown in Figure \ref{fig:6d-8d}f, but that would not affect the
full $n$-vertex requirement.)

A 10-dim or any higher-dimensional masters are too big to be
generated from vector components. Therefore, to obtain the
non-KS MMPH in those dimensions, we rely on minimal KS NBMMPHs
obtained via dimensional upscaling \cite{pw-23} while applying
{\bf M3}. The procedure consists of removing vertices and/or
hyperedges in such a way that an NBMMPH stops being critical,
which enables us to generate smaller critical non-KS NBMMPHs
from it via STATES01.

In Figure \ref{fig:9d-11d}c, we show an 18-9 10-dim critical
obtained via this approach from the 50-15 KS MMPH
master \cite{pw-23}.

In Figure \ref{fig:9d-11d}d, we show a 19-8 11-dim critical
obtained via the same approach from the 50-14 KS MMPH
master \cite{pw-23}.

In the following sections, we stay with this approach while
applying {\bf M3}.

\subsection{\label{subsec:classes12-16}Dimensions 12 to 16}

\begin{figure}[ht]
\begin{center}
  \includegraphics[width=\textwidth]{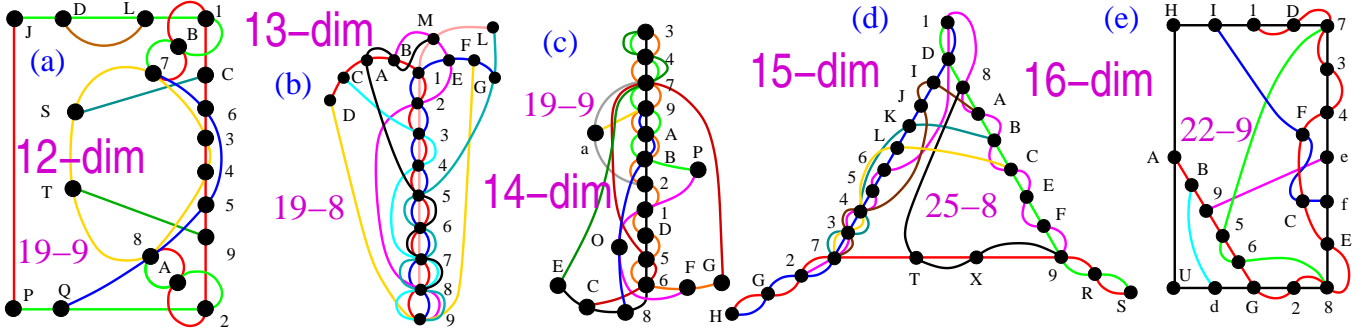}
\end{center}
\caption{(\textbf{a}) Twelve-dim 19-9 critical non-KS
  NBMMPH directly obtained from the master 52-9 via {\bf M3};
  (\textbf{b}) 13-dim 19-8 critical non-KS NBMMPH obtained from the master
  63-16, where the hyperedges do not form any loop with an order of three or higher;
  (\textbf{c}) 14-dim critical obtained from 66-15, where the maximal loop also
  has an order of 2; (\textbf{d}) 15-dim 25-8 critical from the 66-14 master;
  (\textbf{e}) 16-dim 22-9 critical from the 70-9 master, where all criticals are
  obtained via {\bf M3}; all criticals and masters are given in the
  Appendix \ref{app:1i}--\ref{app:1m}.}
\label{fig:12d-16d}
\end{figure}

It has been proven that the minimal complexity (minimal number of
hyperedges or vertices) of the dimensional upscaling of KS MMPHs does
not scale up with dimension \cite{waeg-aravind-pra-17}.
In  \cite{pw-23}, we give a constructive proof that the minimal number
of hyperedges of KS MMPHs repeatedly fluctuates between nine and
sixteen, which confirms this result. In the previous section we
provide constructive generations of critical non-KS
NBMMPHs in dimensions nine to eleven and in this section, in
Fig \ref{fig:12d-16d}(a-e) in dimensions twelve to sixteen, whose
minimal number of hyperedges fluctuates between eight (odd dimensions)
and nine (even dimensions) under the requirement that at least one the
hyperedges contains $n$ vertices, none of which has the multiplicity
$m=1$. In lower dimensions (3--6), the  minimal number of hyperedges
is even smaller. 

\unskip

\section{\label{sec:disc}Discussion}

In this paper, we first generated non-KS contextual NBMMPHs (non-binary MMP
hypergraphs) with the help of master sets generated from
simple vector components whose complexity exponentially scales
with dimension---for dimensions four to eight---and then by means of methods
whose complexity does not scale with dimension. The need
for developing such methods and obtaining MMPHs in higher
dimensions has emerged from recent elaborations of classes of
contextual sets that are not of the KS kind, all of which have
an MMP hypergraph representation. Examples of such elaborations
in the literature and their correspondence with MMPHs are given
in Section \ref{subsec:classes3-5}. In subsequent sections, we
presented generations of non-KS NBMMPHs in spaces of up to 16-dim.

In Section \ref{subsec:form}, we presented the formalism and language of MMPH,
and in Section \ref{subsec:nonks}, we presented the methods of generating them.
In Section \ref{subsec:classes3-5}, we reviewed the most prominent
examples of non-KS sets from the literature in dimensions three to five,
represented them via MMPH formalism, and generated several new
non-KS MMPHs in dimensions four and five with several coordinatizations.
In Section \ref{subsec:classes3-5}, we then went up to the eight-dim
spaces and showed that the arbitrarily exhaustive generation of MMPHs
gets more and more computationally demanding from three-dim to eight-dim
spaces due to the exponentially increasing size of the
MMPH masters obtained from vector components and the exponential
complexity of extracting of NBMMPH classes from them. This is
exacerbated by the ratio of NBMMPHs and BMMPHs, which
starts with less than 0.1\%\ in four-dim spaces and grows exponentially
with the dimension. So, in the nine-dim space in
Section \ref{subsec:classes9-11} with a master containing
9586 vertices and 12,068,705 hyperedges, we can strip any
number of hyperedges from the master, but the probability of
finding any NBMMPH among the obtained MMPHs decreases with
 size (e.g., searching for them in MMPHs with more than a
few thousand hyperedges would take ``forever'' for any
practical purpose). In spaces with dimensions of 10 and greater no
method for obtaining MMPH masters from vector components
is available anymore.

Therefore, to ensure arbitrarily exhaustive generation of MMPHs in
ever higher dimensions, we need a method whose complexity does not
grow with the dimensions. For comparatively small KS MMPHs, such a
method---dimensional upscaling---was recently developed in
 \cite{pw-23} based on previous results in
 \cite{waeg-aravind-pra-17}. In this paper, we put forward a
method of generating non-KS NBMMPHs whose complexity also
does not scale up with the dimensions and which makes use of
KS MMPHs obtained by the former KS method
(in Sections \ref{subsec:classes9-11} and \ref{subsec:classes12-16}).
The method applies to the generation of comparatively small
MMPHs that are still suitable for any practical implementation
since we can always obtain bigger MMPHs at the cost of the time a
generation would take and since really big MMPHs cannot be generated
at all, and even if they could, they would be unimplementable.
The minimal complexity (minimal number of hyperedges or vertices) of
KS MMPHs repeatedly fluctuates between nine and sixteen,
while for non-KS NBMMPHs, it fluctuates between eight (odd
dimensions) and nine (even dimensions) in seven- to sixteen-dim spaces.
In three- to six-dim, it even goes down to three. We provide
a list of them in Table \ref{T:small}.

\begin{table}[ht]
  \caption{The smallest critical non-KS  MMPHs obtained via the small
    vector component method and by the dimensional upscaling method
    via {\bf M1} and {\bf M3}. Notice the steady fluctuation in the
    number of hyperedges over dimensions which is consistent with
    our previous result showing that the minimum complexity of
    NBMMPHs does not grow with the dimensions. The MMPH
    strings and coordinatizations of both the criticals and their
    masters are given in Appendix \ref{app:1}. $\phi$ is the Golden
    ratio, and $\omega$ is the cube root of 1.}

   \setlength{\tabcolsep}{3.3mm}
\resizebox{\linewidth}{!}{\begin{tabular}{cccc}
    \hline
 \textbf{ {dim\ }} & \textbf{Smallest Critical MMPHs} &
  \textbf{Master}  & \textbf{Vector Components}\\
  \hline
  {3-dim}&8-7 (Kochen--Specker's ``bug'')
   &49-36 (Bub's KS MMPH) & $\{0,\pm 1,\pm 2,5\}$\\
  {4-dim}&4-3
  &8-3& $\{0,\pm 1\}$\\
  {4-dim}&16-9
   &58-51& $\{0,\pm\phi,\phi-1\}$\\           
  {5-dim} &7-5& 16-5 & $\{0,\pm 1\}$ \\
  {6-dim}&11-7
   &19-7& $\{0,1,\omega,\omega^2\}$\\
  {7-dim}&14-8
   &34-14& $\{0,\pm 1\}$\\
  {8-dim}&15-9
   &2768-1346016& $\{0,\pm 1\}$\\
  {9-dim} &13-6& 9586-12068705 & $\{0,\pm 1\}$ \\
     {9-dim} &19-8& 47-16 & $\{0,\pm 1\}$ \\
   {10-dim}&18-9& 50-15 & 
   $\{0,\pm 1\}$ \\
   {11-dim}&19-8& 50-14 & 
   $\{0,\pm 1\}$ \\
   {12-dim}&19-9 & 52-9 & 
   $\{0,\pm 1\}$ \\
   {13-dim}&19-8 & 63-16 & 
   $\{0,\pm 1\}$ \\
   {14-dim}&19-9 & 66-15 & 
   $\{0,\pm 1\}$ \\
   {15-dim}&25-8 & 66-14 & 
   $\{0,\pm 1\}$ \\
   {16-dim}&22-9 & 70-9 & 
   $\{0,\pm 1\}$ \\
  \hline  
  \end{tabular}}
\label{T:small}
\end{table}
\unskip

\section{\label{sec:met}Methods}

The methods used to handle quantum contextual sets rely on
algorithms and programs within the MMP language: 
VECFIND, STATES01, MMPSTRIP, MMPSHUFFLE, SUBGRAPH, LOOP, and SHORTD developed in
 \cite{bdm-ndm-mp-1,pmmm05a-corr,pm-ql-l-hql2,pmm-2-10,bdm-ndm-mp-fresl-jmp-10,mfwap-s-11,mp-nm-pka-mw-11,megill-pavicic-mipro-17}.
They are freely available at \url{http://puh.srce.hr/s/Qegixzz2BdjYwFL} 
 MMPHs can be visualized via hypergraph figures consisting of dots
and lines and represented as a string of ASCII characters. The latter representation enables the processing of billions of MMPHs
simultaneously via supercomputers and clusters. For the latter
elaboration, we developed other dynamical programs specifically
 to handle and parallelize jobs with arbitrary numbers of
MMP hypergraph vertices and edges. 

\vspace{6pt} 

\section{\label{sec:concl}Conclusions}

To summarize, based on elaborations of non-KS sets that recently
appeared in the literature and of which we provided several examples
in Section \ref{subsec:classes3-5}, we developed methods of generating
comparatively small non-KS contextual sets in high-dimensional
spaces whose complexity does not grow with the number of dimensions. We provided
examples in all dimensions up to 16. A more detailed summary of
the achieved results is given in Section \ref{sec:disc}.

%%%%%%%%%%%%%%%%%%%%%%%%%%%%%%%%%%%%%%%%%%
\vspace{6pt}

%%%%%%%%%%%%%%%%%%%%%%%%%%%%%%%%%%%%%%%%%%
\begin{acknowledgments}
  Supported by the Ministry of Science and Education of Croatia
  through the Center of Excellence CEMS funding, and by MSE grants
  Nos.~KK.01.1.1.01.0001. Computational support
  provided by the Zagreb University Computing Centre.
  Technical supports of Emir Imamagi\'c from the University of
  Zagreb Computing Centre is gratefully acknowledged.
\end{acknowledgments}

\bigskip 
Programs repository is at {\tt http://puh.srce.hr/s/Qegixzz2BdjYwFL}

\bigskip 
{\bf{Abbreviations.}} The following abbreviations are used in this manuscript:

\noindent 
\begin{tabular}{@{}ll}
  MMPH & McKay--Megill--Pavi\v ci\'c hypergraph
         (Definition \ref{def:mmp})\\
  NBMMPH & Non-binary McKay--Megill--Pavi\v ci\'c hypergraph
         (Definition \ref{df:nbmmph})\\
  BMMPH & Binary McKay--Megill--Pavi\v ci\'c hypergraph
         (Definition \ref{df:bmmph})\\
  KS & Kochen--Specker (Definition \ref{df:ks})\\
  non-KS & Non-Kochen--Specker (Definition \ref{df:nonks})\\
  {\bf M1, M2, M3} & Methods 1,2,3 (Section \ref{subsec:nonks})\\
\end{tabular}

\appendix
\section{\label{app:1} SCII Strings of Non-KS MMPH
     Classes and Their Masters and Supermasters} 

   Below, we give strings and coordinatizations of all MMPHs referred
   to in the main body of the paper. The first hyperedges in a line
   of a critical NBMMPH often correspond to the biggest loops in
   the figures.

\subsection{\label{app:1-3} 3-dim MMPHs}

{\bf 8-7} (KS ``bug'') 123,34,45,567,78,81,26.

\medskip

{\bf 13-7} (filled$\,\,${\bf 8-7}) {\tt 123,394,4A5,567,7B8,8C1,2D6.} {\tt 1}=(0,0,1), {\tt 2}=(0,1,0), {\tt 3}=(1,0,0), {\tt 4}=(0,1,1), {\tt 5}=(1,1,-1), {\tt 6}=(1,0,1), {\tt 7}=(-1,2,1), {\tt 8}=(2,1,0), {\tt 9}=(0,1,-1), {\tt A}=(2,-1,1), {\tt B}=(1,-2,5), {\tt C}=(1,-2,0), {\tt D}=(1,0,-1)

\subsection{\label{app:1a} 4-dim MMPHs}

{\bf 4-3 } {\tt 12,34,1234.}

\medskip

{\bf 8-3} (filled$\,\,${\bf 4-3}) {\tt 1562,3784,1234.} {\tt 1}=(0,0,0,1), {\tt 2}=(0,0,1,0), {\tt 3}=(0,1,0,0), {\tt 4}=(1,0,0,0), {\tt 5}=(1,1,0,0), {\tt 6}=(1,-1,0,0), {\tt 7}=(0,0,1,1), {\tt 8}=(0,0,1,-1)

\medskip

{\bf 16-9}\quad {\tt 3124,49A8,8567,7BC3,DE,FG,GE62,FD51,FECA.}

\medskip

{\bf 20-9} (filled$\,\,${\bf 16-9}) {\tt 1}=(1,0,0,-1), {\tt 2}=(0,1,1,0), {\tt 3}=(1,1,-1,1), {\tt 4}=(1,-1,1,1), {\tt 5}=(1,0,0,1), {\tt 6}=(0,1,-1,0), {\tt 7}=(1,1,1,-1), {\tt 8}=(-1,1,1,1), {\tt 9}=(0,0,1,-1), {\tt A}=(1,1,0,0), {\tt B}=(0,0,1,1), {\tt C}=(1,-1,0,0), {\tt D}=(0,1,0,0), {\tt E}=(0,0,0,1), {\tt F}=(0,0,1,0), {\tt G}=(1,0,0,0), {\tt H}=(1,0,1,0), {\tt I}=(1,0,-1,0), {\tt J}=(0,1,0,1), {\tt K}=(0,1,0,-1)

\medskip

{\bf 24-24} (Peres'$\,\,$super-master) {\tt LMNO,HIJK,DEFG,BCFG,9ADE,78EG,56DF,5678,9ABC,68JK,57HI,ACIK,9BHJ,1234, 4DGO,3EFN,258M,167L,19CM,2ABL,3HKO,4IJN,34NO,12LM.} {\tt 1}=(0,0,0,1), {\tt 2}=(0,0,1,0), {\tt 3}=(1,1,0,0), {\tt 4}=(1,-1,0,0),\break {\tt 5}=(0,1,0,-1), {\tt 6}=(1,0,-1,0), {\tt 7}=(1,0,1,0), {\tt 8}=(0,1,0,1), {\tt 9}=(0,1,-1,0), {\tt A}=(1,0,0,-1), {\tt B}=(1,0,0,1), {\tt C}=(0,1,1,0), {\tt D}=(1,1,1,1), {\tt E}=(1,-1,-1,1), {\tt F}=(1,-1,1,-1), {\tt G}=(1,1,-1,-1), {\tt H}=(-1,1,1,1), {\tt I}=(1,1,-1,1), {\tt J}=(1,1,1,-1), {\tt K}=(1,-1,1,1), {\tt L}=(0,1,0,0),\break  {\tt M}=(1,0,0,0), {\tt N}=(0,0,1,1), {\tt O}=(0,0,1,-1)

\medskip

{\bf 16-9} \qquad {\tt 231,1BC6,654,45GF,FGED,DE32,789A,7BC8,9A.}

\medskip

{\bf 20-9} (filled$\,\,${\bf 16-9}) {\tt 2H31,1BC6,6I54,45GF,FGED,DE32,789A,7BC8,9JKA.} {\tt 1}=(0,0,0,$\phi$), {\tt 2}=($\phi$-1,0,-$\phi$,0),\break  {\tt 3}=($\phi$,0,$\phi$-1,0), {\tt 4}=(0,$\phi$,0,$\phi$), {\tt 5}=(0,$\phi$,0,-$\phi$), {\tt 6}=(0,0,$\phi$,0), {\tt 7}=(0,0,$\phi$,$\phi$), {\tt 8}=(0,0,$\phi$,-$\phi$), {\tt 9}=($\phi$,$\phi$-1,0,0), {\tt A}=($\phi$-1,-$\phi$,0,0), {\tt B}=($\phi$,$\phi$,0,0), {\tt C}=($\phi$,-$\phi$,0,0), {\tt D}=(0,$\phi$-1,0,-$\phi$), {\tt E}=(0,$\phi$,0,$\phi$-1), {\tt F}=($\phi$,0,$\phi$,0), {\tt G}=($\phi$,0,-$\phi$,0), {\tt H}=(0,$\phi$,0,0), {\tt I}=($\phi$,0,0,0),\break  {\tt J}=(0,0,$\phi$,$\phi$-1), {\tt K}=(0,0,$\phi$-1,-$\phi$)

\medskip

{\bf 60-72} (super-master) {\tt 1234,1256,1278,129A,13BC,13DE,13FG,1HI4,1JK4,1LM4,23NO,23PQ,23RS,2TU4,2VW4, 2XY4,Za34,Za56,Za78,Za9A,Z5bc,Zde6,fg34,fg56,fg78,fg9A,hi34,hi56,hi78,hi9A,ajk6,a5lm,ano6,apq6,\break TUBC,TUDE,TUFG,TBbo,TCmd,VWBC,VWDE,VWFG,XYBC,XYDE,XYFG,HINO,HIPQ,HIRS,HNco,HOme,JKNO,JKPQ,JKRS,\break LMNO,LMPQ,LMRS,UBle,UnCc,UrsC,UtCu,jkbc,INld,InOb, IvwO,IxOy,nbco,rsbo,pbcq,vwco,tbuo,xyco,lmde.} {\tt 1}=(0,0,0,$\phi$), {\tt 2}=(0,0,$\phi$,0), {\tt 3}=(0,$\phi$,0,0), {\tt 4}=($\phi$,0,0,0), {\tt 5}=($\phi$,$\phi$,0,0), {\tt 6}=($\phi$,-$\phi$,0,0), {\tt 7}=($\phi$,$\phi$-1,0,0), {\tt 8}=($\phi$-1,-$\phi$,0,0),\break  {\tt 9}=(-$\phi$,$\phi$-1,0,0), {\tt A}=($\phi$-1,$\phi$,0,0), {\tt B}=($\phi$,0,$\phi$,0), {\tt C}=($\phi$,0,-$\phi$,0), {\tt D}=(-$\phi$,0,$\phi$-1,0), {\tt E}=($\phi$-1,0,$\phi$,0), {\tt F}=($\phi$,0,$\phi$-1,0),\break {\tt G}=($\phi$-1,0,-$\phi$,0), {\tt H}=(0,$\phi$,$\phi$,0), {\tt I}=(0,$\phi$,-$\phi$,0), {\tt J}=(0,$\phi$-1,$\phi$,0), {\tt K}=(0,-$\phi$,$\phi$-1,0), {\tt L}=(0,$\phi$-1,-$\phi$,0), {\tt M}=(0,$\phi$,$\phi$-1,0), {\tt N}=($\phi$,0,0,$\phi$), {\tt O}=($\phi$,0,0,-$\phi$), {\tt P}=($\phi$,0,0,$\phi$-1), {\tt Q}=($\phi$-1,0,0,-$\phi$), {\tt R}=(-$\phi$,0,0,$\phi$-1), {\tt S}=($\phi$-1,0,0,$\phi$), {\tt T}=(0,$\phi$,0,$\phi$), {\tt U}=(0,$\phi$,0,-$\phi$), {\tt V}=(0,$\phi$,0,$\phi$-1), {\tt W}=(0,$\phi$-1,0,-$\phi$), {\tt X}=(0,-$\phi$,0,$\phi$-1), {\tt Y}=(0,$\phi$-1,0,$\phi$), {\tt Z}=(0,0,$\phi$,$\phi$), {\tt a}=(0,0,$\phi$,-$\phi$), {\tt b}=($\phi$,-$\phi$,-$\phi$,$\phi$), {\tt c}=($\phi$,-$\phi$,$\phi$,-$\phi$),\break  {\tt d}=($\phi$,$\phi$,$\phi$,-$\phi$), {\tt e}=($\phi$,$\phi$,-$\phi$,$\phi$), {\tt f}=(0,0,$\phi$,$\phi$-1), {\tt g}=(0,0,$\phi$-1,-$\phi$), {\tt h}=(0,0,-$\phi$,$\phi$-1), {\tt i}=(0,0,$\phi$-1,$\phi$), {\tt j}=($\phi$,$\phi$,$\phi$-1,$\phi$-1),\break  {\tt k}=($\phi$-1,$\phi$-1,-$\phi$,-$\phi$),{\tt l}=(-$\phi$,$\phi$,$\phi$,$\phi$), {\tt m}=($\phi$,-$\phi$,$\phi$,$\phi$), {\tt n}=($\phi$,$\phi$,$\phi$,$\phi$), {\tt o}=($\phi$,$\phi$,-$\phi$,-$\phi$), {\tt p}=(-$\phi$,-$\phi$,$\phi$-1,$\phi$-1), {\tt q}=($\phi$-1,$\phi$-1,$\phi$,$\phi$),\break  {\tt r}=($\phi$,$\phi$-1,$\phi$,$\phi$-1), {\tt s}=($\phi$-1,-$\phi$,$\phi$-1,-$\phi$), {\tt t}=(-$\phi$,$\phi$-1,-$\phi$,$\phi$-1), {\tt u}=($\phi$-1,$\phi$,$\phi$-1,$\phi$), {\tt v}=($\phi$,$\phi$-1,$\phi$-1,$\phi$), {\tt w}=($\phi$-1,-$\phi$,-$\phi$,$\phi$-1),\break  {\tt x}=(-$\phi$,$\phi$-1,$\phi$-1,-$\phi$), {\tt y}=($\phi$-1,$\phi$,$\phi$,$\phi$-1)

\subsection{\label{app:1b} 5-dim MMPHs}

{\bf 7-5} {\tt 41235,56,674,234,714.}

\medskip

{\bf 16-5} ({\bf 7-5}$\,\,$filled) {\tt 41235,589A6,6BC74,2DE34,7FG14.} {\tt 1}=(0,0,1,0,0), {\tt 2}=(1,-1,0,0,0), {\tt 3}=(1,1,0,0,0), {\tt 4}=(0,0,0,0,1), {\tt 5}=(0,0,0,1,0), {\tt 6}=(0,1,1,0,0), {\tt 7}=(1,0,0,1,0), {\tt 8}=(1,0,0,0,-1), {\tt 9}=(0,1,-1,0,0), {\tt A}=(1,0,0,0,1), {\tt B}=(1,-1,1,-1,0),\break  {\tt C}=(1,1,-1,-1,0), {\tt D}=(0,0,1,1,0), {\tt E}=(0,0,1,-1,0), {\tt F}=(0,1,0,0,0), {\tt G}=(1,0,0,-1,0)

\medskip

{\bf 16-9} \quad {\tt 63457,75B9E,EFGD,DC,CA86,12345,89125,AB,FGE.} 

\medskip

{\bf 26-9} ({\bf 16-9}$\,\,$filled) {\tt 63457,75B9E,EFHGD,DIJKC,CAL86,12345,89125,AMNOB,FPQGE.} {\tt 1}=(1,1,1,-1,0), {\tt 2}=(1,1,-1,1,0), {\tt 3}=(1,-1,1,1,0), {\tt 4}=(-1,1,1,1,0), {\tt 5}=(0,0,0,0,1), {\tt 6}=(0,0,1,-1,0), {\tt 7}=(1,1,0,0,0), {\tt 8}=(0,0,1,1,0), {\tt 9}=(1,-1,0,0,0), {\tt A}=(0,1,0,0,1), {\tt B}=(0,0,1,0,0), {\tt C}=(1,0,0,0,0), {\tt D}=(0,1,1,0,0), {\tt E}=(0,0,0,1,0), {\tt F}=(1,0,0,0,1), {\tt G}=(0,1,-1,0,0), {\tt H}=(1,0,0,0,-1), {\tt I}=(0,0,0,1,1), {\tt J}=(0,1,-1,1,-1), {\tt K}=(0,1,-1,-1,1), {\tt L}=(0,1,0,0,-1), {\tt M}=(1,1,0,-1,-1), {\tt N}=(1,-1,0,-1,1), {\tt O}=(1,0,0,1,0), {\tt P}=(1,1,1,0,-1),\break  {\tt Q}=(-1,1,1,0,1)

\medskip

{\bf 105-136} (super-master) {\tt 12345,12367,12489,12AB5,134CD,13EF5,1GH45,1GH67,1G6IJ,1GKL7,1H6MN,1HOP7,\break 1EF89,1E8IP,1E9KN,1F8ML,1F9OJ,1ABCD,1ACJP,1ADLN,1QRST,1QUVW,1XYSZ,1XabW,1BCMK,1BDOI,1cYVd,1caeT,\break 1fRbd,1fUeZ,1OIJP,1MKLN,234gh,23ij5,2kl45,2kl67,2k6mn,2kop7,2l6qr,2lst7,2ij89,2i8mt,2i9or,2j8qp,\break 2j9sn,2ABgh,2Agtn,2Ahpr,2uvSw,2uxyW,2z!S",2z\#\$W,2Bgoq,2Bhsm,2\%!yd,2\%\#ew,2\&v\$d,2\&xe",2smtn,2oqpr,\break '(345,'(367,'(489,'(AB5,'36)*,'3-/7,'48:;,'4<=9,'A>?5,'@[B5,(36\textbackslash ],(3\^{}\_7,(48`\{,(4|\}9,(A$\sim$+15,\break (+2+3B5,3ijCD,3iC\_*,3iD-\textbackslash , 3jC/],3jD\^{}),3EFgh,3Eg\_),3Eh/\textbackslash ,3+4+5Vw,3+4+6yT,3+7+8V",3+7+9\$T,3Fg-],\break 3Fh\^{}*,3+A+8yZ,3+A+9bw,3+B+5\$Z,3+B+6b",3\^{}\_)*,3-/\textbackslash ],kl4CD,klEF5,k4C\};,k4D<`,kE+3?5,kF@$\sim$5,l4C=\{,\break l4D|:,lE[+15,lF+2>5,GH4gh,GHij5,G4g\}:,G4h=`,Gi+3>5,Gj[$\sim$5,+C4+5Rx,+C4+6vU,+C+7X\%5,+C+Azc5,+D4+8R\#,\break +D4+9!U,+D+4X\&5,+D+Buc5,H4g<\{,H4h|;,Hi@+15,Hj+2?5,+E4+8va,+E4+9Yx,+E+4zf5,+E+BQ\%5,+F4+5!a,\break +F4+6Y\#,+F+7uf5,+F+AQ\&5,4|\}:;,4<=`\{,+2+3>?5,@[$\sim$+15.} {\tt 1}=(0,0,0,0,1), {\tt 2}=(0,0,0,1,0), {\tt '}=(0,0,0,1,1),\break  {\tt (}=(0,0,0,1,-1), {\tt 3}=(0,0,1,0,0), {\tt k}=(0,0,1,0,1), {\tt l}=(0,0,1,0,-1), {\tt G}=(0,0,1,1,0), {\tt +C}=(0,0,1,1,1), {\tt +D}=(0,0,1,1,-1),\break  {\tt H}=(0,0,1,-1,0), {\tt +E}=(0,0,1,-1,1), {\tt +F}=(0,0,-1,1,1), {\tt 4}=(0,1,0,0,0), {\tt i}=(0,1,0,0,1), {\tt j}=(0,1,0,0,-1), {\tt E}=(0,1,0,1,0),\break  {\tt +4}=(0,1,0,1,1), {\tt +7}=(0,1,0,1,-1), {\tt F}=(0,1,0,-1,0), {\tt +A}=(0,1,0,-1,1), {\tt +B}=(0,-1,0,1,1), {\tt A}=(0,1,1,0,0), {\tt u}=(0,1,1,0,1),\break  {\tt z}=(0,1,1,0,-1), {\tt Q}=(0,1,1,1,0), {\tt +2}=(0,1,1,1,1), {\tt @}=(0,1,1,1,-1), {\tt X}=(0,1,1,-1,0), {\tt [}=(0,1,1,-1,1), {\tt +3}=(0,1,1,-1,-1),\break  {\tt B}=(0,1,-1,0,0), {\tt \%}=(0,1,-1,0,1), {\tt \&}=(0,-1,1,0,1), {\tt c}=(0,1,-1,1,0), {\tt $\sim$}=(0,1,-1,1,1), {\tt >}=(0,1,-1,1,-1), {\tt f}=(0,-1,1,1,0),\break  {\tt ?}=(0,1,-1,-1,1), {\tt +1}=(0,-1,1,1,1), {\tt 5}=(1,0,0,0,0), {\tt g}=(1,0,0,0,1), {\tt h}=(1,0,0,0,-1), {\tt C}=(1,0,0,1,0), {\tt +8}=(1,0,0,1,1),\break  {\tt +5}=(1,0,0,1,-1), {\tt D}=(1,0,0,-1,0), {\tt +6}=(1,0,0,-1,1), {\tt +9}=(-1,0,0,1,1), {\tt 8}=(1,0,1,0,0), {\tt !}=(1,0,1,0,1), {\tt v}=(1,0,1,0,-1),\break  {\tt Y}=(1,0,1,1,0), {\tt |}=(1,0,1,1,1), {\tt <}=(1,0,1,1,-1), {\tt R}=(1,0,1,-1,0), {\tt =}=(1,0,1,-1,1), {\tt \}}=(1,0,1,-1,-1), {\tt 9}=(1,0,-1,0,0),\break  {\tt x}=(1,0,-1,0,1), {\tt \#}=(-1,0,1,0,1), {\tt U}=(1,0,-1,1,0), {\tt `}=(1,0,-1,1,1), {\tt :}=(1,0,-1,1,-1), {\tt a}=(-1,0,1,1,0), {\tt ;}=(1,0,-1,-1,1),\break  {\tt \{}=(-1,0,1,1,1), {\tt 6}=(1,1,0,0,0), {\tt \$}=(1,1,0,0,1), {\tt y}=(1,1,0,0,-1), {\tt b}=(1,1,0,1,0), {\tt \^{}}=(1,1,0,1,1), {\tt -}=(1,1,0,1,-1),\break  {\tt V}=(1,1,0,-1,0), {\tt /}=(1,1,0,-1,1), {\tt \_}=(1,1,0,-1,-1), {\tt e}=(1,1,1,0,0), {\tt s}=(1,1,1,0,1), {\tt o}=(1,1,1,0,-1), {\tt O}=(1,1,1,1,0), {\tt M}=(-1,1,1,1,0), {\tt I}=(1,-1,-1,1,0), {\tt K}=(1,1,1,-1,0), {\tt q}=(-1,1,1,0,1), {\tt m}=(1,-1,-1,0,1), {\tt S}=(1,1,-1,0,0), {\tt p}=(1,1,-1,0,1), {\tt t}=(1,1,-1,0,-1),\break  {\tt L}=(1,1,-1,1,0), {\tt d}=(-1,1,1,0,0), {\tt J}=(1,-1,1,-1,0), {\tt P}=(1,1,-1,-1,0), {\tt N}=(1,-1,1,1,0), {\tt n}=(1,-1,1,0,-1), {\tt 7}=(1,-1,0,0,0),\break  {\tt w}=(1,-1,0,0,1), {\tt "}=(-1,1,0,0,1), {\tt T}=(1,-1,0,1,0), {\tt \textbackslash }=(1,-1,0,1,1), {\tt )}=(1,-1,0,1,-1), {\tt Z}=(-1,1,0,1,0), {\tt *}=(1,-1,0,-1,1),\break  {\tt ]}=(-1,1,0,1,1),  {\tt W}=(1,-1,1,0,0), {\tt r}=(1,-1,1,0,1)

\medskip

{\bf 10-9} \cite[Fig. 5(a)]{cabello-svozil-18}
{\tt 17835,27846,91,97,92,8A,34,6A,5A.}

\medskip

{\bf 31-9} ({\bf 10-9}$\,\,$filled) {\tt 17835,27846,9BCD1,9EFG7,9HIJ2,8KLMA,3NOP4,6QRSA,5TUVA.} {\tt 1}=(0,0,1,0,-1), {\tt 2}=(0,0,0,1,-1), {\tt 3}=(0,0,0,1,0), {\tt 4}=(0,0,1,0,0), {\tt 5}=(0,0,1,0,1), {\tt 6}=(0,0,0,1,1), {\tt 7}=(0,1,0,0,0), {\tt 8}=(1,0,0,0,0), {\tt 9}=(0,0,1,1,1),\break  {\tt A}=(0,1,1,1,-1), {\tt B}=(0,0,-1,2,-1), {\tt C}=(1,2,0,0,0), {\tt D}=(2,-1,0,0,0), {\tt E}=(0,0,-1,-1,2), {\tt F}=(1,0,1,-1,0), {\tt G}=(2,0,-1,1,0),\break  {\tt H}=(0,0,2,-1,-1), {\tt I}=(1,1,0,0,0), {\tt J}=(1,-1,0,0,0), {\tt K}=(0,0,1,-1,0), {\tt L}=(0,1,0,0,1), {\tt M}=(0,-1,1,1,1), {\tt N}=(0,0,0,0,1), {\tt O}=(-1,2,0,0,0), {\tt P}=(2,1,0,0,0), {\tt Q}=(0,1,1,-1,1), {\tt R}=(1,1,-1,0,0), {\tt S}=(2,-1,1,0,0), {\tt T}=(0,1,0,-1,0), {\tt U}=(2,1,-1,1,1), {\tt V}=(2,-1,1,-1,-1)

\subsection{\label{app:1c} 6-dim MMPHs}

{\bf 19-7} \qquad {\tt 7!8gw,woO6i,i;)EB,B<:b7,'!)Jb6,'o:8Eu,;<OJgu.}

\medskip

{\bf 11-7} \qquad {\tt w!g,wi6,'!)Jb6,'u,B)i,Jgu,Bb.}

\medskip

{\bf 21-7} ({\bf 11-7,$\,\,$19-7}$\,\,$filled) {\tt w!78\#g,woOi6\%,'!)Jb6,'o:8Eu,;B)i\#E,;$<$OJgu,B$<$7:b\%.} {\tt w}=($\omega$,1,1,$\omega^2$,1,$\omega$),\break  {\tt '}=($\omega^2$,$\omega$,$\omega$,1,1,1), {\tt ;}=($\omega$,1,1,$\omega^2$,$\omega$,1), {\tt !}=($\omega$,1,$\omega^2$,1,$\omega$,1), {\tt B}=(1,$\omega^2$,$\omega$,1,$\omega$,1), {\tt $<$}=($\omega$,1,$\omega$,1,1,$\omega^2$), {\tt o}=(1,$\omega^2$,1,1,1,$\omega^2$),\break  {\tt )}=($\omega$,$\omega^2$,1,1,1,$\omega$), {\tt 7}=(1,$\omega$,$\omega^2$,1,1,$\omega$), {\tt O}=(1,$\omega$,$\omega$,$\omega^2$,1,1), {\tt :}=($\omega^2$,1,1,1,$\omega$,$\omega$), {\tt 8}=(1,$\omega$,1,$\omega^2$,$\omega$,1), {\tt i}=($\omega^2$,1,$\omega^2$,1,1,1),\break  {\tt J}=(1,$\omega$,1,1,$\omega$,$\omega^2$), {\tt \#}=(1,1,1,1,$\omega^2$,$\omega^2$), {\tt b}=(1,1,1,$\omega$,1,1), {\tt 6}=(1,1,$\omega$,1,$\omega^2$,$\omega$), {\tt g}=($\omega^2$,$\omega^2$,1,1,1,1), {\tt E}=(1,1,$\omega$,$\omega^2$,1,$\omega$),\break  {\tt \%}=($\omega$,$\omega$,1,1,$\omega^2$,1), {\tt u}=(1,1,$\omega^2$,1,$\omega^2$,1)

\medskip

{\bf 79-162} (\%\ and \#\ stripped off of {\bf 81-162}) {\tt 123456,12789A,1BCD5E,1B7FGH,1ICJ9K,1I3LGM,1NODAP,1NQ4HR, 1SOJ6T,1SU8MR,1VQLET,1VUFKP,WXY45Z,WX7abA,Wcde5E,Wc7Ffg,WIdJbh,WIYifM,WjkeAP,WjQ4gl,WSkJZm,WSnaMl, WoQiEm,WonFhP,pqY89Z,pq3ab6,pcre9K,pc3Lsg,pBrDbh,pBYisH,pjte6T,pjU8gu,pNtDZm,pNnaHu,pvnLhT,pvUiKm, wxyD5Z,wx7azH,w!"e56,w!78g,wI"Lzh,wIyiK,w\$keHR,w\$ODgl,wVkLZ,wV\&aKl,woOi6,wo\&8hR,'(yDbA,'(Y4zH,\break '!)Jb6,'!Y8*M,'c)LzE,'cyF*K,'-kJHu,'-tDMl, 'vkLA/,'v:4Kl,'otF6/,'o:8Eu,;("e9A,;(34g,;x)J9Z,;x3a*M,\break ;B)iE,;B"F*h,;$<$teMR,;$<$OJgu,;vOiA=,;v$>$4hR,;VtFZ=,;V$>$aEu,?!reGM,?!CJsg,?qyFGZ,?qCazE,?2yisA,?2r4zh,\break ?\$:eET,?\$UFg/,?-\&JhT,?-UiM,?N:4Z,?N\&aA/,@(deGH,@(CDfg,@X)LGZ,@XCa*K,@2)if6,@2d8*h,@$<$:eKP,@$<$QLg/,\break  @-$>$DhP,@-QiH=,@S:8Z=,@S$>$a6/,[xdJsH,[xrDfM,[X"LsA,[Xr4K,[q"Ff6,[qd8E,[$<$\&JKm,[$<$nLM, [\$$>$DEm,[\$nFH=,\break [j$>$46,[j\&8A=,(S"Uzm,(SynT,(VdUb,(VY\&fT,(oCn9,(o3\&Gm,xj)UzP,xjyQ*T, xvdU5/,xv7:fT,xorQ9/,xo3:sP,\break !N)nP,!N"Q*m,!vCn5=,!v7$>$Gm,!VrQb=,!VY$>$sP,X\$)UbR,X\$YO*T, X-"U5u,X-7tT,XorOGu,XoCtsR,q$<$yQbR,q$<$YOzP,\break q-"Q9l,q-3kP,qvdOGl,qvCkfR,2$<$yn5u,2$<$7tzm, 2\$)n9l,2\$3k*m,2Vdtsl,2Vrkfu,c-3\&5=,c-7$>$9,cN)\&fR,cNdO*,\break cSy$>$sR,cSrOz=,B$<$Y\&5/,B$<$7:b,Bj)\&Gl,BjCk*,BS":sl,BSrk/,I\$Y$>$9/,I\$3:b=,Ijy$>$Gu,IjCtz=,IN":fu,INdt/.}

\medskip

{\bf 81-162} {\tt 123456,12789A,1BCD5E,1B7FGH,1ICJ9K,1I3LGM,1NODAP,1NQ4HR,1SOJ6T,1SU8MR,1VQLET,1VUFKP,\break WXY45Z,WX7abA,Wcde5E,Wc7Ffg,WIdJbh,WIYifM,WjkeAP,WjQ4gl,WSkJZm,WSnaMl,WoQiEm,WonFhP,pqY89Z,pq3ab6,\break pcre9K,pc3Lsg,pBrDbh,pBYisH,pjte6T,pjU8gu,pNtDZm,pNnaHu, pvnLhT,pvUiKm,wxyD5Z,wx7azH,w!"e56,\break w!78\#g,wI"Lzh,wIyi\#K,w\$keHR,w\$ODgl,wVkLZ\%,wV\&aKl, woOi6\%,wo\&8hR,'(yDbA,'(Y4zH,'!)Jb6,'!Y8*M,\break 'c)LzE,'cyF*K,'-kJHu,'-tDMl,'vkLA/,'v:4Kl,'otF6/,'o:8Eu,;("e9A,;(34\#g,;x)J9Z,;x3a*M,;B)i\#E,\break ;B"F*h,;$<$teMR,;$<$OJgu,;vOiA=,;v$>$4hR,;VtFZ=,;V$>$aEu,?!reGM,?!CJsg,?qyFGZ,?qCazE,?2yisA,?2r4zh,\break ?\$:eET,?\$UFg/,?-\&JhT,?-UiM\%,?N:4Z\%,?N\&aA/,@(deGH,@(CDfg,@X)LGZ,@XCa*K,@2)if6,@2d8*h,@$<$:eKP,\break @$<$QLg/,@-$>$DhP, @-QiH=,@S:8Z=,@S$>$a6/,[xdJsH,[xrDfM,[X"LsA,[Xr4\#K,[q"Ff6,[qd8\#E,[$<$\&JKm,[$<$nLM\%,\break [\$$>$DEm,[\$nFH=,[j$>$46\%,[j\&8A=,(S"Uzm,(Syn\#T,(VdUb\%,(VY\&fT,(oCn9\%,(o3\&Gm,xj)UzP,xjyQ*T,xvdU5/,\break xv7:fT,xorQ9/,xo3:sP,!N)n\#P,!N"Q*m,!vCn5=,!v7$>$Gm,!VrQb=,!VY$>$sP,X\$)UbR,X\$YO*T,X-"U5u,X-7t\#T,\break XorOGu,XoCtsR,q$<$yQbR,q$<$YOzP,q-"Q9l,q-3k\#P,qvdOGl,qvCkfR,2$<$yn5u, 2$<$7tzm,2\$)n9l,2\$3k*m,2Vdtsl,\break 2Vrkfu,c-3\&5=,c-7$>$9\%,cN)\&fR,cNdO*\%,cSy$>$sR,cSrOz=,B$<$Y\&5/,B$<$7:b\%,Bj)\&Gl,BjCk*\%,BS":sl,BSrk\#/,\break I\$Y$>$9/,I\$3:b=,Ijy$>$Gu,IjCtz=,IN":fu, INdt\#/.} {\tt 1}=($\omega$,1,1,1,1,1), {\tt W}=($\omega$,1,1,1,$\omega^2$,$\omega$), {\tt p}=($\omega$,1,1,1,$\omega$,$\omega^2$),\break  {\tt w}=($\omega$,1,1,$\omega^2$,1,$\omega$), {\tt '}=($\omega^2$,$\omega$,$\omega$,1,1,1), {\tt ;}=($\omega$,1,1,$\omega^2$,$\omega$,1), {\tt ?}=($\omega$,1,1,$\omega$,1,$\omega^2$), {\tt @}=($\omega$,1,1,$\omega$,$\omega^2$,1), {\tt [}=(1,$\omega^2$,$\omega^2$,1,1,1),\break  {\tt (}=($\omega$,1,$\omega^2$,1,1,$\omega$), {\tt x}=($\omega^2$,$\omega$,1,$\omega$,1,1), {\tt !}=($\omega$,1,$\omega^2$,1,$\omega$,1), {\tt X}=($\omega^2$,$\omega$,1,1,$\omega$,1), {\tt q}=($\omega^2$,$\omega$,1,1,1,$\omega$), {\tt 2}=(1,$\omega^2$,$\omega$,$\omega$,1,1),\break  {\tt c}=($\omega$,1,$\omega^2$,$\omega$,1,1), {\tt B}=(1,$\omega^2$,$\omega$,1,$\omega$,1), {\tt I}=(1,$\omega^2$,$\omega$,1,1,$\omega$), {\tt $<$}=($\omega$,1,$\omega$,1,1,$\omega^2$), {\tt \$}=($\omega$,1,$\omega$,1,$\omega^2$,1), {\tt -}=(1,$\omega^2$,1,$\omega^2$,1,1),\break  {\tt j}=($\omega$,1,$\omega$,$\omega^2$,1,1), {\tt N}=(1,$\omega^2$,1,$\omega$,$\omega$,1), {\tt S}=(1,$\omega^2$,1,$\omega$,1,$\omega$), {\tt v}=(1,$\omega^2$,1,1,$\omega^2$,1), {\tt V}=(1,$\omega^2$,1,1,$\omega$,$\omega$), {\tt o}=(1,$\omega^2$,1,1,1,$\omega^2$),\break  {\tt )}=($\omega$,$\omega^2$,1,1,1,$\omega$), {\tt "}=($\omega^2$,1,$\omega$,$\omega$,1,1), {\tt y}=($\omega$,$\omega^2$,1,1,$\omega$,1), {\tt d}=($\omega^2$,1,$\omega$,1,$\omega$,1), {\tt r}=($\omega^2$,1,$\omega$,1,1,$\omega$), {\tt C}=(1,$\omega$,$\omega^2$,$\omega$,1,1),\break  {\tt Y}=($\omega$,$\omega^2$,1,$\omega$,1,1), {\tt 3}=(1,$\omega$,$\omega^2$,1,$\omega$,1), {\tt 7}=(1,$\omega$,$\omega^2$,1,1,$\omega$), {\tt k}=($\omega^2$,1,1,$\omega$,$\omega$,1), {\tt t}=($\omega^2$,1,1,$\omega$,1,$\omega$), {\tt O}=(1,$\omega$,$\omega$,$\omega^2$,1,1),\break  {\tt :}=($\omega^2$,1,1,1,$\omega$,$\omega$), {\tt $>$}=($\omega^2$,1,1,1,1,$\omega^2$), {\tt \&}=($\omega^2$,1,1,1,$\omega^2$,1), {\tt Q}=(1,$\omega$,$\omega$,1,$\omega^2$,1), {\tt n}=($\omega^2$,1,1,$\omega^2$,1,1), {\tt U}=(1,$\omega$,$\omega$,1,1,$\omega^2$),\break  {\tt a}=($\omega$,$\omega^2$,$\omega$,1,1,1), {\tt 8}=(1,$\omega$,1,$\omega^2$,$\omega$,1), {\tt 4}=(1,$\omega$,1,$\omega^2$,1,$\omega$), {\tt F}=(1,$\omega$,1,$\omega$,$\omega^2$,1), {\tt i}=($\omega^2$,1,$\omega^2$,1,1,1), {\tt L}=(1,$\omega$,1,$\omega$,1,$\omega^2$),\break  {\tt D}=(1,$\omega$,1,1,$\omega^2$,$\omega$), {\tt J}=(1,$\omega$,1,1,$\omega$,$\omega^2$), {\tt *}=(1,1,1,1,1,$\omega$), {\tt z}=(1,1,1,1,$\omega$,1), {\tt \#}=(1,1,1,1,$\omega^2$,$\omega^2$), {\tt e}=(1,$\omega$,1,1,1,1),\break  {\tt b}=(1,1,1,$\omega$,1,1), {\tt 5}=(1,1,1,$\omega$,$\omega$,$\omega^2$), {\tt 9}=(1,1,1,$\omega$,$\omega^2$,$\omega$), {\tt f}=(1,1,1,$\omega^2$,1,$\omega^2$), {\tt G}=(1,1,1,$\omega^2$,$\omega$,$\omega$), {\tt s}=(1,1,1,$\omega^2$,$\omega^2$,1),\break  {\tt Z}=(1,1,$\omega$,1,1,1), {\tt A}=(1,1,$\omega$,1,$\omega$,$\omega^2$), {\tt 6}=(1,1,$\omega$,1,$\omega^2$,$\omega$), {\tt H}=(1,1,$\omega$,$\omega$,1,$\omega^2$), {\tt g}=($\omega^2$,$\omega^2$,1,1,1,1), {\tt M}=(1,1,$\omega$,$\omega$,$\omega^2$,1),\break  {\tt E}=(1,1,$\omega$,$\omega^2$,1,$\omega$), {\tt K}=(1,1,$\omega$,$\omega^2$,$\omega$,1), {\tt h}=($\omega$,$\omega$,$\omega^2$,1,1,1), {\tt =}=($\omega$,$\omega$,1,1,1,$\omega^2$), {\tt \%}=($\omega$,$\omega$,1,1,$\omega^2$,1), {\tt l}=(1,1,$\omega^2$,1,1,$\omega^2$),\break  {\tt R}=(1,1,$\omega^2$,1,$\omega$,$\omega$), {\tt u}=(1,1,$\omega^2$,1,$\omega^2$,1), {\tt /}=(1,1,$\omega^2$,$\omega^2$,1,1), \quad {\tt m}=($\omega$,$\omega$,1,$\omega^2$,1,1), \ \quad {\tt P}=(1,1,$\omega^2$,$\omega$,1,$\omega$), \quad\ {\tt T}=(1,1,$\omega^2$,$\omega$,$\omega$,1)

\medskip

{\bf 31-16} {\tt 237,7HG,GTUVRP,PRNQSM,MIJKL2,235,7235,3NOKLM,WXJRSM,WYZGRV,aOQbcM,aYdHce,XIfbcM,fgUHce,\break gTUGH,YZdGH.}

\medskip

{\bf 44-16} (master for {\bf 31-16}) {\tt 123456,72389A,723B5C,7DEFGH,2IJKLM,3NOKLM,PNQRSM,PTUGRV,WXJRSM,WYZGRV,\break aOQbcM,aYdHce,XIfbcM,fgUHce,gTUhGH,iYZdGH.} {\tt 1}=(0,0,1,-1,1,0), {\tt 7}=(0,0,-1,1,1,0), {\tt 2}=(0,1,0,0,0,1), {\tt 3}=(0,1,0,0,0,-1), {\tt P}=(0,1,0,0,1,0), {\tt W}=(0,1,0,0,-1,0), {\tt a}=(0,1,0,1,0,0), {\tt X}=(0,1,0,1,1,1), {\tt D}=(0,1,0,1,-1,0), {\tt N}=(0,1,0,1,-1,1), {\tt I}=(0,1,0,1,-1,-1), {\tt f}=(0,1,0,-1,0,0), {\tt O}=(0,1,0,-1,1,1), {\tt J}=(0,1,0,-1,1,-1), {\tt Q}=(0,-1,0,1,1,1), {\tt E}=(0,1,1,0,1,0), {\tt g}=(0,1,1,1,1,0),\break  {\tt i}=(0,1,1,1,-1,0), {\tt Y}=(0,1,1,-1,1,0), {\tt T}=(0,1,1,-1,-1,0), {\tt Z}=(0,1,-1,1,1,0), {\tt U}=(0,1,-1,1,-1,0), {\tt F}=(0,-1,1,1,0,0),\break  {\tt h}=(0,1,-1,-1,1,0), {\tt d}=(0,-1,1,1,1,0), {\tt G}=(1,0,0,0,0,1), {\tt H}=(1,0,0,0,0,-1), {\tt 8}=(1,0,0,1,-1,0), {\tt B}=(1,0,0,-1,1,0), {\tt 4}=(-1,0,0,1,1,0), {\tt 9}=(1,0,1,0,1,0), {\tt b}=(1,0,1,0,1,-1), {\tt c}=(1,0,1,0,-1,1), {\tt 5}=(1,0,1,1,0,0), {\tt R}=(1,0,1,1,0,-1), {\tt K}=(1,0,1,1,1,0), {\tt S}=(1,0,1,-1,0,1), {\tt L}=(1,0,1,-1,-1,0), {\tt M}=(1,0,-1,0,0,0), {\tt 6}=(1,0,-1,0,1,0), {\tt e}=(1,0,-1,0,1,1), {\tt C}=(-1,0,1,0,1,0), {\tt A}=(-1,0,1,1,0,0), {\tt V}=(-1,0,1,1,0,1)

\medskip

{\bf 117-116} (strip-master of 236-1216 super-master) {\tt 123456,123789,1AB4CD,1AB7EF,1GHI5F,1GHJC9,1KLI8D,\break 1KLJE6,MN8COP,MNF6QR,STE5OP,ST9DQR,UVWXYZ,UVabcd,UefgYZ,Uefabh,Uijgcd,UijWXh,UBk4XZ,UBk7ac,U3l4bd,\break U3l7WY,ULmIXY,ULmJbc,UHnIad,UHnJWZ,opVgYZ,opVabh,opijgh,opabqR,opYZPr,oistuv,oiwxyz,oj!"\#\$,oj\%\&'(,\break )*Vgcd,)*VWXh,)*efgh,)*WXqR,)*cdPr,)e-/uv,)e:;yz,)f!"$<$=,)f\%\&$>$?,*e@[\#\$,*e\textbackslash ]'(,*fst\^{}\textunderscore,*fwx`\{,pi@[$<$=,\break pi\textbackslash ]$>$?,pj-/\^{}\textunderscore,pj:;`\{,2|V4XZ,2|V7ac,2|3l47,2|ac\}$\sim$,2|XZ+1+2,2389u\{,2356\^{}z,2l\textbackslash "+3+4,2l\%[+5+6,A+7V4bd,\break A+7V7WY,A+7Bk47,A+7WY\}$\sim$,A+7bd+1+2,ABEFu\{,ABCD\^{}z,Ak\textbackslash "+8+9,Ak\%[+A+B,G+CVIXY,G+CVJbc,G+CHnIJ,\break G+Cbc+D+E,G+CXY+F+G,GHC9\#?,GH5F$<$(,Gn:t+3+B,Gnw/+8+6,K+HVIad,K+HVJWZ,K+HLmIJ,K+HWZ+D+E,K+Had+F+G,\break KLE6\#?,KL8D$<$(,Km:t+5+9,Kmw/+A+4,+7B@\&+3+4,+7B!]+5+6,+7k89y\textunderscore,+7k56`v,|3@\&+8+9,|3!]+A+B,\break |lEFy\textunderscore,|lCD`v,+HL-x+3+B,+HLs;+8+6,+HmC9'=,+Hm5F$>$\$,+CH-x+5+9,+CHs;+A+4,+CnE6'=,+Cn8D$>$\$,efab+IO,\break efYZQ+J,ijWX+IO,ijcdQ+J,Bkac+K+L,BkXZ+M+N,3lWY+K+L,3lbd+M+N,Lmbc+O+P,LmXY+Q+R,HnWZ+O+P,Hnad+Q+R.} {\tt 1}=(0,1,0,-1,0,0), {\tt M}=(0,1,0,-1,1,1), {\tt S}=(0,1,0,-1,1,-1), {\tt T}=(0,1,0,-1,-1,1), {\tt N}=(0,-1,0,1,1,1), {\tt U}=(0,1,1,0,0,0), {\tt o}=(0,1,1,0,1,1), {\tt )}=(0,1,1,0,1,-1), {\tt *}=(0,1,1,0,-1,1), {\tt p}=(0,1,1,0,-1,-1), {\tt 2}=(0,1,1,1,0,1), {\tt A}=(0,1,1,1,0,-1), {\tt G}=(0,1,1,1,1,0),\break  {\tt K}=(0,1,1,1,-1,0), {\tt +7}=(0,1,1,-1,0,1), {\tt |}=(0,1,1,-1,0,-1), {\tt +H}=(0,1,1,-1,1,0), {\tt +C}=(0,1,1,-1,-1,0), {\tt V}=(0,1,-1,0,0,0),\break  {\tt e}=(0,1,-1,0,1,1), {\tt i}=(0,1,-1,0,1,-1), {\tt j}=(0,1,-1,0,-1,1), {\tt f}=(0,-1,1,0,1,1), {\tt B}=(0,1,-1,1,0,1), {\tt 3}=(0,1,-1,1,0,-1),\break  {\tt L}=(0,1,-1,1,1,0), {\tt H}=(0,1,-1,1,-1,0), {\tt l}=(0,1,-1,-1,0,1), {\tt k}=(0,-1,1,1,0,1), {\tt n}=(0,1,-1,-1,1,0), {\tt m}=(0,-1,1,1,1,0), {\tt I}=(1,0,0,0,0,1), {\tt J}=(1,0,0,0,0,-1), {\tt 4}=(1,0,0,0,1,0), {\tt 7}=(1,0,0,0,-1,0), {\tt g}=(1,0,0,1,0,0), {\tt W}=(1,0,0,1,1,1), {\tt a}=(1,0,0,1,1,-1), {\tt b}=(1,0,0,1,-1,1), {\tt X}=(1,0,0,1,-1,-1), {\tt h}=(1,0,0,-1,0,0), {\tt c}=(1,0,0,-1,1,1), {\tt Y}=(1,0,0,-1,1,-1), {\tt Z}=(1,0,0,-1,-1,1), {\tt d}=(-1,0,0,1,1,1), {\tt E}=(1,0,1,0,1,1), {\tt 8}=(1,0,1,0,1,-1), {\tt C}=(1,0,1,0,-1,1), {\tt 5}=(1,0,1,0,-1,-1), {\tt -}=(1,0,1,1,0,1), {\tt s}=(1,0,1,1,0,-1), {\tt @}=(1,0,1,1,1,0), {\tt !}=(1,0,1,1,-1,0), {\tt :}=(1,0,1,-1,0,1), {\tt w}=(1,0,1,-1,0,-1), {\tt \textbackslash }=(1,0,1,-1,1,0), {\tt \%}=(1,0,1,-1,-1,0), {\tt 9}=(1,0,-1,0,1,1), {\tt F}=(1,0,-1,0,1,-1),\break  {\tt 6}=(1,0,-1,0,-1,1), {\tt D}=(-1,0,1,0,1,1), {\tt t}=(1,0,-1,1,0,1), {\tt /}=(1,0,-1,1,0,-1), {\tt "}=(1,0,-1,1,1,0), {\tt [}=(1,0,-1,1,-1,0),\break  {\tt x}=(1,0,-1,-1,0,1), {\tt ;}=(-1,0,1,1,0,1), {\tt \&}=(1,0,-1,-1,1,0), {\tt ]}=(-1,0,1,1,1,0), {\tt +A}=(1,1,0,0,1,1), {\tt +5}=(1,1,0,0,1,-1),\break  {\tt +8}=(1,1,0,0,-1,1), {\tt +3}=(1,1,0,0,-1,-1), {\tt $>$}=(1,1,0,1,0,1), {\tt '}=(1,1,0,1,0,-1), {\tt `}=(1,1,0,1,1,0), {\tt y}=(1,1,0,1,-1,0),\break  {\tt $<$}=(1,1,0,-1,0,1), {\tt \# }=(1,1,0,-1,0,-1), {\tt \^{}}=(1,1,0,-1,1,0), {\tt u}=(1,1,0,-1,-1,0), {\tt +Q}=(1,1,1,0,0,1), {\tt +O}=(1,1,1,0,0,-1),\break  {\tt +M}=(1,1,1,0,1,0), {\tt +K}=(1,1,1,0,-1,0), {\tt Q}=(1,1,1,1,0,0), {\tt +I}=(1,1,1,-1,0,0), {\tt O}=(-1,1,1,1,0,0), {\tt +F}=(1,1,-1,0,0,1),\break  {\tt +D}=(1,1,-1,0,0,-1), {\tt +1}=(1,1,-1,0,1,0), {\tt \}}=(1,1,-1,0,-1,0), {\tt P}=(1,1,-1,1,0,0), {\tt +J}=(1,-1,-1,1,0,0), {\tt q}=(1,1,-1,-1,0,0),\break  {\tt +L}=(-1,1,1,0,1,0), {\tt +N}=(1,-1,-1,0,1,0), {\tt +6}=(1,-1,0,0,1,1), {\tt +B}=(1,-1,0,0,1,-1), {\tt +4}=(1,-1,0,0,-1,1), {\tt +9}=(-1,1,0,0,1,1),\break  {\tt (}=(1,-1,0,1,0,1), {\tt ?}=(1,-1,0,1,0,-1), {\tt z}=(1,-1,0,1,1,0), {\tt \{}=(1,-1,0,1,-1,0), {\tt \$}=(1,-1,0,-1,0,1), {\tt =}=(-1,1,0,1,0,1),\break  {\tt v}=(1,-1,0,-1,1,0), {\tt \textunderscore}=(-1,1,0,1,1,0), {\tt +G}=(1,-1,1,0,0,1), {\tt +E}=(1,-1,1,0,0,-1), {\tt +2}=(1,-1,1,0,1,0), {\tt $\sim$}=(1,-1,1,0,-1,0),\break  {\tt r}=(1,-1,1,1,0,0), {\tt +P}=(-1,1,1,0,0,1), {\tt +R}=(1,-1,-1,0,0,1), {\tt R}=(1,-1,1,-1,0,0)

\subsection{\label{app:1d} 7-dim MMPHs}

{\bf 14-8}\quad {\tt 12567,189A5BC,189DE7,189HJ,189HB,2D,2EC,AJ.}

\medskip

{\bf 31-13} {\tt 1234,189DEF,189GHIJ,189KHBL,2MNDOIP,2MNEOCL,2MNGKF,QRNSAJP,QT4U,RTV9,WXMS,WYV8AJP,XY3U.}
                 
\medskip

{\bf 34-14} (master$\,\,$for$\,\,${\bf 14-8}$\,\,$and$\,\,${\bf 31-13} ) {\tt 1234567,189A5BC,189DE7F,189GHIJ,189KHBL,2MNDOIP,2MNEOCL,2MNGK6F,\break QRNSAJP,QT4U567,RTV9567,WXMS567,WYV8AJP,XY3U567.} {\tt 1}=(0,0,0,1,0,0,0); {\tt 2}=(0,0,1,0,0,0,0); {\tt 3}=(1,-1,0,0,0,0,0); {\tt 4}=(1,1,0,0,0,0,0); {\tt 5}=(0,0,0,0,0,0,1); {\tt 6}=(0,0,0,0,1,1,0); {\tt 7}=(0,0,0,0,1,-1,0); {\tt 8}=(0,1,-1,0,0,0,0); {\tt 9}=(0,1,1,0,0,0,0);\break  {\tt A}=(0,0,0,0,1,0,0); {\tt B}=(1,0,0,0,0,-1,0); {\tt C}=(1,0,0,0,0,1,0); {\tt D}=(1,0,0,0,1,1,-1); {\tt E}=(-1,0,0,0,1,1,1); {\tt F}=(1,0,0,0,0,0,1);\break  {\tt G}=(1,0,0,0,1,-1,-1); {\tt H}=(1,0,0,0,1,1,1); {\tt I}=(1,0,0,0,-1,0,0); {\tt J}=(0,0,0,0,0,1,-1); {\tt K}=(1,0,0,0,-1,1,-1); {\tt L}=(0,0,0,0,1,0,-1); {\tt M}=(0,1,0,1,0,0,0); {\tt N}=(0,1,0,-1,0,0,0); {\tt O}=(1,0,0,0,1,-1,1); {\tt P}=(0,0,0,0,0,1,1); {\tt Q}=(-1,1,1,1,0,0,0); {\tt R}=(1,1,-1,1,0,0,0);\break  {\tt S}=(1,0,1,0,0,0,0); {\tt T}=(1,-1,1,1,0,0,0); {\tt U}=(0,0,1,-1,0,0,0); {\tt V}=(1,0,0,-1,0,0,0); {\tt W}=(1,-1,-1,1,0,0,0); {\tt X}=(1,1,-1,-1,0,0,0);\break  {\tt Y}=(1,1,1,1,0,0,0).

\subsection{\label{app:1e} 8-dim MMPHs}

{\bf 15-9} {\tt 17426538,8E,E2M,M3N,N4O,O5U,U7Q,Q6H,H1.}

\medskip

{\bf 36-9} (master for {\bf 15-9}) {\tt 17426538,8ABDF9CE,ELaR2YVM,M3WSDKZN,NCJXR4TO,OV5PSBIU,UZ9GXa7Q,QTY6PWAH,\break HIKFGJL1.} {\tt 1}=(0,0,0,0,0,0,0,1), {\tt 2}=(0,0,0,0,0,0,1,0), {\tt 3}=(0,0,0,0,0,1,0,0), {\tt 4}=(0,0,0,0,1,0,0,0), {\tt 5}=(0,0,1,1,0,0,0,0),\break  {\tt 6}=(0,0,-1,1,0,0,0,0), {\tt 7}=(1,1,0,0,0,0,0,0), {\tt 8}=(-1,1,0,0,0,0,0,0), {\tt 9}=(0,0,0,0,0,0,1,1), {\tt A}=(0,0,1,1,1,-1,0,0),\break  {\tt B}=(1,1,0,0,0,0,-1,1), {\tt C}=(1,1,0,0,0,0,1,-1), {\tt D}=(0,0,-1,0,1,0,0,0), {\tt E}=(0,0,0,1,0,1,0,0), {\tt F}=(0,0,1,-1,1,1,0,0),\break  {\tt G}=(0,0,0,1,1,0,0,0), {\tt H}=(0,0,1,1,-1,1,0,0), {\tt I}=(1,0,0,0,0,0,1,0), {\tt J}=(0,0,-1,0,0,1,0,0), {\tt K}=(-1,0,0,0,0,0,1,0),\break  {\tt L}=(0,1,0,0,0,0,0,0), {\tt M}=(0,0,1,0,1,0,0,0), {\tt N}=(0,0,0,1,0,0,0,0), {\tt O}=(-1,1,0,0,0,0,1,1), {\tt P}=(0,0,0,0,1,1,0,0),\break  {\tt Q}=(-1,1,0,0,0,0,1,-1), {\tt R}=(1,0,0,0,0,0,0,1), {\tt S}=(0,-1,0,0,0,0,0,1), {\tt T}=(0,-1,0,0,0,0,1,0), {\tt U}=(0,0,-1,1,-1,1,0,0),\break  {\tt V}=(0,0,-1,1,1,-1,0,0), {\tt W}=(1,1,0,0,0,0,1,1), {\tt X}=(0,0,1,0,0,1,0,0), {\tt Y}=(-1,0,0,0,0,0,0,1), {\tt Z}=(-1,1,0,0,0,0,-1,1),\break  {\tt a}=(0,0,-1,-1,1,1,0,0)

\subsection{\label{app:1f} 9-dim MMPHs}

{\bf 13-6} \quad {\tt SU,1G42U,1S,472acefhK,G72,4U.}

\medskip 
{\bf 44-6} ({\bf 13-6} filled) \quad {\tt SUCDEFOQ6,1G42U8H95,1SAIMSVXbi,472acefhK,G72LWYZdg,4UBJ3NPRT.} {\tt 1}=(0,0,0,0,0,0,0,1,0); {\tt 2}=(0,0,0,0,0,0,1,0,0); {\tt 3}=(0,0,0,0,1,1,0,0,0); {\tt 4}=(0,0,0,1,0,0,0,0,1); {\tt 5}=(0,0,0,1,0,0,0,0,-1);  {\tt 6}=(0,0,0,1,0,0,-1,1,0);\break  {\tt 7}=(0,0,1,0,0,0,0,1,0); {\tt 8}=(0,0,1,0,0,1,0,0,0); {\tt 9}=(0,0,1,0,0,-1,0,0,0); {\tt A}=(0,0,1,0,-1,0,1,0,0);  {\tt B}=(0,0,1,0,-1,1,1,0,0);\break  {\tt C}=(0,0,-1,-1,1,1,0,1,1); {\tt D}=(0,0,1,-1,-1,-1,0,1,1); {\tt H}=(0,1,0,0,1,0,0,0,0); {\tt E}=(0,1,0,0,1,-1,1,1,-1);\break   {\tt F}=(0,1,0,0,1,-1,-1,-1,1); {\tt G}=(0,1,0,0,-1,0,0,0,0); {\tt I}=(0,1,0,1,1,1,1,0,1); {\tt J}=(0,1,0,-1,0,0,0,-1,1);\break  {\tt K}=(1,-1,1,1,-1,1,0,-1,-1);  {\tt L}=(0,1,0,-1,1,1,0,0,0); {\tt M}=(0,1,0,-1,1,-1,1,0,-1); {\tt N}=(0,1,-1,0,0,0,1,1,0); {\tt O}=(0,1,1,1,0,1,1,0,1); {\tt P}=(0,1,1,1,1,-1,1,-1,-1);  {\tt Q}=(0,1,1,-1,0,1,-1,0,-1); {\tt R}=(0,1,-1,0,0,0,1,1,0); {\tt S}=(0,-1,1,0,1,0,0,0,0); {\tt U}=(1,0,0,0,0,0,0,0,0); {\tt V}=(1,0,0,0,0,-1,0,0,1);  {\tt W}=(1,0,0,1,0,1,0,0,1); {\tt X}=(1,0,0,-1,0,1,0,0,0); {\tt Y}=(1,0,0,-1,0,-1,0,0,1); {\tt Z}=(1,0,1,0,0,0,0,-1,-1); {\tt a}=(1,1,0,0,-1,-1,0,0,0); {\tt b}=(1,1,1,1,0,0,-1,0,-1); {\tt c}=(1,1,1,-1,1,1,0,-1,1); {\tt d}=(-1,1,1,1,1,-1,0,-1,1); {\tt e}=(1,-1,-1,-1,-1,1,0,1,1); {\tt f}=(1,1,-1,1,1,1,0,1,-1); {\tt g}=(1,1,-1,1,1,-1,0,1,-1); {\tt h}=(1,-1,0,0,1,-1,0,0,0); {\tt i}=(1,-1,-1,1,0,0,1,0,-1). 

\medskip

{\bf 19-8} \quad {\tt 1234567,129ABCDE,13FGH5IJE,GB6E,AJE,97E,LH4DE,FIE.}

\medskip

{\bf 47-16} (master for {\bf 19-8}) {\tt 123456789,12ABCDEF,13HIJ5KL,1AMCLNOPQ,1BRSETUVQ,1H4567VWX,1ICDEFYOX,\break 1ZJ4FTUW9,1ZCDEFYP8,23abRSET,cdIeMD6f,cgaZMCLN,dghA4567,ijhBRk7f,ilbZJ4FT,jlHeSkKN.}\break  {\tt 1}=(1,0,0,0,0,0,0,0,0), {\tt 2}=(0,1,0,0,0,0,0,0,0), {\tt 3}=(0,0,1,0,0,0,0,0,0), {\tt c}=(1,1,1,1,0,0,0,0,0), {\tt d}=(1,-1,1,-1,0,0,0,0,0),\break  {\tt g}=(1,-1,-1,1,0,0,0,0,0), {\tt i}=(1,-1,-1,-1,0,0,0,0,0), {\tt j}=(1,-1,1,1,0,0,0,0,0), {\tt l}=(1,1,1,-1,0,0,0,0,0), {\tt h}=(1,1,0,0,0,0,0,0,0),\break  {\tt A}=(0,0,1,1,0,0,0,0,0), {\tt B}=(0,0,1,-1,0,0,0,0,0), {\tt H}=(0,1,0,1,0,0,0,0,0), {\tt I}=(0,1,0,-1,0,0,0,0,0), {\tt e}=(1,0,-1,0,0,0,0,0,0),\break  {\tt a}=(1,0,0,-1,0,0,0,0,0), {\tt b}=(1,0,0,1,0,0,0,0,0), {\tt Z}=(0,1,-1,0,0,0,0,0,0), {\tt J}=(0,0,0,0,1,0,0,0,0), {\tt 4}=(0,0,0,0,0,1,0,0,0),\break  {\tt 5}=(0,0,0,0,0,0,1,0,0), {\tt M}=(0,0,0,0,1,1,1,1,0), {\tt C}=(0,0,0,0,1,-1,1,-1,0), {\tt D}=(0,0,0,0,1,-1,-1,1,0), {\tt R}=(0,0,0,0,1,-1,-1,-1,0),\break  {\tt S}=(0,0,0,0,1,-1,1,1,0), {\tt k}=(0,0,0,0,1,1,1,-1,0), {\tt E}=(0,0,0,0,1,1,0,0,0), {\tt F}=(0,0,0,0,0,0,1,1,0), {\tt T}=(0,0,0,0,0,0,1,-1,0),\break  {\tt K}=(0,0,0,0,0,1,0,1,0), {\tt L}=(0,0,0,0,0,1,0,-1,0), {\tt N}=(0,0,0,0,1,0,-1,0,0), {\tt 6}=(0,0,0,0,1,0,0,-1,0), {\tt 7}=(0,0,0,0,1,0,0,1,0),\break  {\tt f}=(0,0,0,0,0,1,-1,0,0), {\tt Y}=(0,1,1,1,0,0,0,0,1), {\tt O}=(0,1,-1,1,0,0,0,0,-1), {\tt P}=(0,1,1,-1,0,0,0,0,-1), {\tt U}=(0,1,1,1,0,0,0,0,-1),\break  {\tt V}=(0,1,-1,-1,0,0,0,0,-1), {\tt W}=(0,1,1,-1,0,0,0,0,1), {\tt Q}=(0,1,0,0,0,0,0,0,1), {\tt X}=(0,0,1,0,0,0,0,0,-1), {\tt 8}=(0,0,0,1,0,0,0,0,-1),\break  {\tt 9}=(0,0,0,1,0,0,0,0,1)

\subsection{\label{app:1g} 10-dim MMPHs}

{\bf 18-9} \quad {\tt 1BC5DEFGH9,1BCKL9A,T5DEU,TPR,TbL9A,CbKP,bKG,bKFR,bKUHA.}

\medskip

{\bf 50-15}  (master for {\bf 18-9}) {\tt 12BCUVfgik,1DEJXYceoq,1DELMVWajk,1DEMNRSblm,1DEOPRTajk,1DEOQYZajk,\break 2GHLNXZajk,2GHPQUWblm,45EFUVcdmn,46GIJSTajk,46GIUVceoq,56ABUVdeij, 78ACUVfhpq,79HIUVabop,\break 89DFUVghln.} {\tt 1}=(1,0,0,0,0,0,0,0,0,0), {\tt 2}=(0,1,0,0,0,0,0,0,0,0), {\tt 3}=(0,0,1,0,0,0,0,0,0,0), {\tt 4}=(1,1,1,1,0,0,0,0,0,0),\break  {\tt 5}=(1,-1,1,-1,0,0,0,0,0,0), {\tt 6}=(1,-1,-1,1,0,0,0,0,0,0), {\tt 7}=(1,-1,-1,-1,0,0,0,0,0,0), {\tt 8}=(1,-1,1,1,0,0,0,0,0,0),\break  {\tt 9}=(1,1,1,-1,0,0,0,0,0,0), {\tt A}=(1,1,0,0,0,0,0,0,0,0), {\tt B}=(0,0,1,1,0,0,0,0,0,0), {\tt C}=(0,0,1,-1,0,0,0,0,0,0), {\tt D}=(0,1,0,1,0,0,0,0,0,0), {\tt E}=(0,1,0,-1,0,0,0,0,0,0), {\tt F}=(1,0,-1,0,0,0,0,0,0,0), {\tt G}=(1,0,0,-1,0,0,0,0,0,0), {\tt H}=(1,0,0,1,0,0,0,0,0,0), {\tt I}=(0,1,-1,0,0,0,0,0,0,0), {\tt J}=(0,0,0,0,1,0,0,0,0,0), {\tt K}=(0,0,0,0,0,1,0,0,0,0), {\tt L}=(0,0,1,0,1,1,1,0,0,0), {\tt M}=(0,0,1,0,-1,1,-1,0,0,0), {\tt N}=(0,0,1,0,-1,-1,1,0,0,0), {\tt O}=(0,0,1,0,-1,-1,-1,0,0,0), {\tt P}=(0,0,1,0,-1,1,1,0,0,0), {\tt Q}=(0,0,1,0,1,1,-1,0,0,0), {\tt R}=(0,0,1,0,1,0,0,0,0,0), {\tt S}=(0,0,0,0,0,1,1,0,0,0), {\tt T}=(0,0,0,0,0,1,-1,0,0,0), {\tt U}=(0,0,0,0,1,0,1,0,0,0), {\tt V}=(0,0,0,0,1,0,-1,0,0,0), {\tt W}=(0,0,1,0,0,-1,0,0,0,0), {\tt X}=(0,0,1,0,0,0,-1,0,0,0), {\tt Y}=(0,0,1,0,0,0,1,0,0,0), {\tt Z}=(0,0,0,0,1,-1,0,0,0,0), {\tt a}=(0,0,0,0,0,0,0,1,0,0), {\tt b}=(0,0,0,0,0,0,0,0,1,0), {\tt c}=(0,0,0,0,0,1,0,1,1,1), {\tt d}=(0,0,0,0,0,1,0,-1,1,-1), {\tt e}=(0,0,0,0,0,1,0,-1,-1,1), {\tt f}=(0,0,0,0,0,1,0,-1,-1,-1), {\tt g}=(0,0,0,0,0,1,0,-1,1,1),\break  {\tt h}=(0,0,0,0,0,1,0,1,1,-1), {\tt i}=(0,0,0,0,0,1,0,1,0,0), {\tt j}=(0,0,0,0,0,0,0,0,1,1), {\tt k}=(0,0,0,0,0,0,0,0,1,-1), {\tt l}=(0,0,0,0,0,0,0,1,0,1),\break  {\tt m}=(0,0,0,0,0,0,0,1,0,-1),$\,${\tt n}=(0,0,0,0,0,1,0,0,-1,0),$\,${\tt o}=(0,0,0,0,0,1,0,0,0,-1),$\,${\tt p}=(0,0,0,0,0,1,0,0,0,1),$\,${\tt q}=(0,0,0,0,0,0,0,1,-1,0)

\subsection{\label{app:1h} 11-dim MMPHs}

{\bf 19-8}\quad{\tt 123456789AB,1234567CDF,1GHKLMDA,27KL9,567MC,567B,H8F,G8F.}

\medskip

{\bf 50-14} (master for {\bf 19-8}) {\tt 123456789AB,1234567CDEF,1GHIJKLMDAN,1GHIJKLOPQR,27STUVKL8FW,27STUVKL9QX,\break 347YZabMDAN,567cdefMCXR,567cdefgOEW,567cdefgPBN,cdhijkJV8FW,eYlmnIUk9QX, fZoHTjmn8FW,abGShilo8FW.} {\tt 1}=(0,0,1,1,1,1,0,0,0,0,0), {\tt 2}=(0,0,1,-1,1,-1,0,0,0,0,0), {\tt 3}=(0,0,0,1,0,-1,0,0,0,0,0), {\tt 4}=(0,0,1,0,-1,0,0,0,0,0,0),\break  {\tt 5}=(0,1,0,0,0,0,0,0,0,0,0), {\tt 6}=(1,0,0,0,0,0,0,0,0,0,0), {\tt 7}=(0,0,0,0,0,0,1,0,0,0,0), {\tt 8}=(0,0,0,1,0,0,0,0,0,0,0),\break  {\tt 9}=(0,0,1,0,0,0,0,0,0,0,0), {\tt A}=(0,0,0,0,0,1,0,0,0,0,0), {\tt B}=(0,0,0,0,1,0,0,0,0,0,0), {\tt C}=(1,-1,1,0,1,0,0,0,0,0,0),\break  {\tt D}=(1,1,0,1,0,1,0,0,0,0,0), {\tt E}=(1,1,0,-1,0,-1,0,0,0,0,0), {\tt F}=(-1,1,1,0,1,0,0,0,0,0,0), {\tt G}=(0,1,-1,1,0,0,1,0,0,0,0),\break  {\tt H}=(1,0,1,1,0,0,0,-1,0,0,0), {\tt I}=(1,0,0,0,1,1,0,1,0,0,0), {\tt J}=(0,1,0,0,-1,1,-1,0,0,0,0), {\tt K}=(0,0,1,0,-1,0,1,1,0,0,0),\break  {\tt L}=(0,0,0,1,0,-1,-1,1,0,0,0), {\tt M}=(1,0,1,0,0,-1,1,0,0,0,0), {\tt N}=(0,-1,1,0,0,1,0,1,0,0,0), {\tt O}=(-1,1,0,0,0,0,1,1,0,0,0),\break  {\tt P}=(1,0,-1,-1,0,0,0,1,0,0,0), {\tt Q}=(0,1,1,-1,0,0,-1,0,0,0,0), {\tt R}=(1,0,0,1,-1,0,-1,0,0,0,0), {\tt S}=(0,1,0,1,1,0,0,1,0,0,0),\break  {\tt T}=(1,1,0,0,0,0,1,-1,0,0,0), {\tt U}=(0,1,0,0,1,-1,-1,0,0,0,0), {\tt V}=(1,0,0,0,-1,-1,0,1,0,0,0), {\tt W}=(1,1,0,-1,0,1,0,0,0,0,0),\break  {\tt X}=(1,-1,-1,0,1,0,0,0,0,0,0), {\tt Y}=(0,0,0,0,0,0,0,0,1,0,0), {\tt Z}=(0,0,0,0,0,0,0,0,0,1,0), {\tt a}=(0,0,0,0,0,0,0,1,1,1,1),\break  {\tt b}=(0,0,0,0,0,0,0,1,-1,1,-1), {\tt c}=(0,0,0,0,0,0,0,1,-1,-1,1), {\tt d}=(0,0,0,0,0,0,0,1,-1,-1,-1), {\tt e}=(0,0,0,0,0,0,0,1,-1,1,1),\break  {\tt f}=(0,0,0,0,0,0,0,1,1,1,-1), {\tt g}=(0,0,0,0,0,0,0,1,1,0,0), {\tt h}=(0,0,0,0,0,0,0,0,0,1,1), {\tt i}=(0,0,0,0,0,0,0,0,0,1,-1),\break  {\tt j}=(0,0,0,0,0,0,0,0,1,0,1), {\tt k}=(0,0,0,0,0,0,0,0,1,0,-1), {\tt l}=(0,0,0,0,0,0,0,1,0,-1,0), {\tt m}=(0,0,0,0,0,0,0,1,0,0,-1),\break  {\tt n}=(0,0,0,0,0,0,0,1,0,0,1),\qquad {\tt o}=(0,0,0,0,0,0,0,0,1,-1,0)

\subsection{\label{app:1i} 12-dim MMPHs}

{\bf 19-9}\quad{\tt 123456789ABC,17DJBL,28PQA,PJ,3478ST,5678Q,SC,T9,DL}

\medskip

{\bf 52-9} (master for {\bf 19-9}) {\tt 123456789ABC,17DEFGHIJBKL,28MNOPHIQAKR,3478STUVWXYR,5678ZabcdWQe,\break ZafghiGPjJke,bSlmnFOijdoC,cTpENhmn9qYo,UVDMfglpkXqL.} {\tt 1}=(0,0,1,1,1,1,0,0,0,0,0,0), {\tt 2}=(0,0,1,-1,1,-1,0,0,0,0,0,0), {\tt 3}=(0,0,0,1,0,-1,0,0,0,0,0,0), {\tt 4}=(0,0,1,0,-1,0,0,0,0,0,0,0), {\tt 5}=(0,1,0,0,0,0,0,0,0,0,0,0), {\tt 6}=(1,0,0,0,0,0,0,0,0,0,0,0),\break  {\tt 7}=(0,0,0,0,0,0,0,1,0,0,0,0), {\tt 8}=(0,0,0,0,0,0,1,0,0,0,0,0), {\tt Z}=(0,0,0,1,0,0,0,0,0,0,0,0), {\tt a}=(0,0,1,0,0,0,0,0,0,0,0,0),\break  {\tt b}=(0,0,0,0,0,1,0,0,0,0,0,0), {\tt c}=(0,0,0,0,1,0,0,0,0,0,0,0), {\tt S}=(1,-1,1,0,1,0,0,0,0,0,0,0), {\tt T}=(1,1,0,1,0,1,0,0,0,0,0,0),\break  {\tt U}=(1,1,0,-1,0,-1,0,0,0,0,0,0), {\tt V}=(-1,1,1,0,1,0,0,0,0,0,0,0), {\tt D}=(0,1,-1,1,0,0,1,0,0,0,0,0), {\tt M}=(1,0,1,1,0,0,0,-1,0,0,0,0),\break  {\tt f}=(1,0,0,0,1,1,0,1,0,0,0,0), {\tt g}=(0,1,0,0,-1,1,-1,0,0,0,0,0), {\tt l}=(0,0,1,0,-1,0,1,1,0,0,0,0), {\tt p}=(0,0,0,1,0,-1,-1,1,0,0,0,0),\break  {\tt E}=(1,0,1,0,0,-1,1,0,0,0,0,0), {\tt N}=(0,-1,1,0,0,1,0,1,0,0,0,0), {\tt h}=(-1,1,0,0,0,0,1,1,0,0,0,0), {\tt m}=(1,0,-1,-1,0,0,0,1,0,0,0,0),\break  {\tt n}=(0,1,1,-1,0,0,-1,0,0,0,0,0), {\tt F}=(1,0,0,1,-1,0,-1,0,0,0,0,0), {\tt O}=(0,1,0,1,1,0,0,1,0,0,0,0), {\tt i}=(1,1,0,0,0,0,1,-1,0,0,0,0),\break  {\tt G}=(0,1,0,0,1,-1,-1,0,0,0,0,0), {\tt P}=(1,0,0,0,-1,-1,0,1,0,0,0,0), {\tt H}=(1,1,0,-1,0,1,0,0,0,0,0,0), {\tt I}=(1,-1,-1,0,1,0,0,0,0,0,0,0),\break  {\tt j}=(0,0,0,0,0,0,0,0,1,0,0,0), {\tt d}=(0,0,0,0,0,0,0,0,0,1,0,0), {\tt J}=(0,0,0,0,0,0,0,0,0,1,1,0), {\tt k}=(0,0,0,0,0,0,0,0,0,1,-1,0),\break  {\tt W}=(0,0,0,0,0,0,0,0,1,0,1,0), {\tt Q}=(0,0,0,0,0,0,0,0,1,0,-1,0), {\tt 9}=(0,0,0,0,0,0,0,0,1,-1,0,0), {\tt e}=(0,0,0,0,0,0,0,0,0,0,0,1),\break  {\tt A}=(0,0,0,0,0,0,0,0,1,1,1,1), {\tt B}=(0,0,0,0,0,0,0,0,1,1,-1,-1), {\tt K}=(0,0,0,0,0,0,0,0,1,-1,1,-1), {\tt X}=(0,0,0,0,0,0,0,0,1,-1,-1,-1),\break  {\tt q}=(0,0,0,0,0,0,0,0,1,1,1,-1), {\tt Y}=(0,0,0,0,0,0,0,0,1,1,-1,1), {\tt L}=(0,0,0,0,0,0,0,0,1,0,0,1), {\tt o}=(0,0,0,0,0,0,0,0,0,0,1,1),\break  {\tt C}=(0,0,0,0,0,0,0,0,0,0,1,-1),\quad {\tt R}=(0,0,0,0,0,0,0,0,0,1,0,-1) 

\subsection{\label{app:1j} 13-dim MMPHs}

{\bf 19-8} \quad {\tt 123456789ABCD,123456789EFG,12345678ILM,289EBM,34789C,56789LG,5678ABM,9FD.}

\medskip 
  
{\bf 63-16} (master for {\bf 19-8}) {\tt 123456789ABCD,123456789EFGH,12345678IJKLM,17NOPQRSTUVWM,28XYZaRS9EbBM,\break 3478cdef9bghH,3478cdef9WhiC,3478cdefIjVkM,5678lmno9LpGq,5678lmnoATrBM,lmstuvQa9WpFD,lmstuvQaEKwxM,\break ncyz!PZv9AkLM,od"OYuz!9kgiq,od"OYuz!bUjxM,efNXsty"rJwWM}. {1}=(0,0,1,1,1,1,0,0,0,0,0,0,0),\break  {\tt 2}=(0,0,1,-1,1,-1,0,0,0,0,0,0,0), {\tt 3}=(0,0,0,1,0,-1,0,0,0,0,0,0,0), {\tt 4}=(0,0,1,0,-1,0,0,0,0,0,0,0,0), {\tt 5}=(0,1,0,0,0,0,0,0,0,0,0,0,0), {\tt 6}=(1,0,0,0,0,0,0,0,0,0,0,0,0), {\tt 7}=(0,0,0,0,0,0,0,1,0,0,0,0,0), {\tt 8}=(0,0,0,0,0,0,1,0,0,0,0,0,0), {\tt l}=(0,0,0,1,0,0,0,0,0,0,0,0,0), {\tt m}=(0,0,1,0,0,0,0,0,0,0,0,0,0), {\tt n}=(0,0,0,0,0,1,0,0,0,0,0,0,0), {\tt o}=(0,0,0,0,1,0,0,0,0,0,0,0,0), {\tt c}=(1,-1,1,0,1,0,0,0,0,0,0,0,0), {\tt d}=(1,1,0,1,0,1,0,0,0,0,0,0,0), {\tt e}=(1,1,0,-1,0,-1,0,0,0,0,0,0,0), {\tt f}=(-1,1,1,0,1,0,0,0,0,0,0,0,0), {\tt N}=(0,1,-1,1,0,0,1,0,0,0,0,0,0), {\tt X}=(1,0,1,1,0,0,0,-1,0,0,0,0,0), {\tt s}=(1,0,0,0,1,1,0,1,0,0,0,0,0), {\tt t}=(0,1,0,0,-1,1,-1,0,0,0,0,0,0), {\tt y}=(0,0,1,0,-1,0,1,1,0,0,0,0,0), {\tt "}=(0,0,0,1,0,-1,-1,1,0,0,0,0,0), {\tt O}=(1,0,1,0,0,-1,1,0,0,0,0,0,0), {\tt Y}=(0,-1,1,0,0,1,0,1,0,0,0,0,0), {\tt u}=(-1,1,0,0,0,0,1,1,0,0,0,0,0), {\tt z}=(1,0,-1,-1,0,0,0,1,0,0,0,0,0), {\tt !}=(0,1,1,-1,0,0,-1,0,0,0,0,0,0), {\tt P}=(1,0,0,1,-1,0,-1,0,0,0,0,0,0), {\tt Z}=(0,1,0,1,1,0,0,1,0,0,0,0,0), {\tt v}=(1,1,0,0,0,0,1,-1,0,0,0,0,0), {\tt Q}=(0,1,0,0,1,-1,-1,0,0,0,0,0,0), {\tt a}=(1,0,0,0,-1,-1,0,1,0,0,0,0,0), {\tt R}=(1,1,0,-1,0,1,0,0,0,0,0,0,0), {\tt S}=(1,-1,-1,0,1,0,0,0,0,0,0,0,0), {\tt 9}=(0,0,0,0,0,0,0,0,1,0,0,0,0), {\tt A}=(0,0,0,0,0,0,0,0,0,1,0,0,0), {\tt E}=(0,0,0,0,0,0,0,0,0,1,1,0,0), {\tt b}=(0,0,0,0,0,0,0,0,0,1,-1,0,0), {\tt T}=(0,0,0,0,0,0,0,0,1,0,1,0,0), {\tt r}=(0,0,0,0,0,0,0,0,1,0,-1,0,0), {\tt I}=(0,0,0,0,0,0,0,0,1,-1,0,0,0), {\tt B}=(0,0,0,0,0,0,0,0,0,0,0,1,0), {\tt J}=(0,0,0,0,0,0,0,0,1,1,1,1,0), {\tt K}=(0,0,0,0,0,0,0,0,1,1,-1,-1,0), {\tt w}=(0,0,0,0,0,0,0,0,1,-1,1,-1,0), {\tt U}=(0,0,0,0,0,0,0,0,1,-1,-1,-1,0), {\tt j}=(0,0,0,0,0,0,0,0,1,1,1,-1,0), {\tt V}=(0,0,0,0,0,0,0,0,1,1,-1,1,0), {\tt x}=(0,0,0,0,0,0,0,0,1,0,0,1,0), {\tt k}=(0,0,0,0,0,0,0,0,0,0,1,1,0), {\tt L}=(0,0,0,0,0,0,0,0,0,0,1,-1,0), {\tt W}=(0,0,0,0,0,0,0,0,0,1,0,-1,0), {\tt M}=(0,0,0,0,0,0,0,0,0,0,0,0,1), {\tt p}=(0,0,0,0,0,0,0,0,0,1,1,1,1),$\,${\tt F}=(0,0,0,0,0,0,0,0,0,1,-1,1,-1),$\,${\tt G}=(0,0,0,0,0,0,0,0,0,1,-1,-1,1),$\,${\tt g}=(0,0,0,0,0,0,0,0,0,1,1,-1,1),\break  {\tt h}=(0,0,0,0,0,0,0,0,0,1,1,1,-1), {\tt i}=(0,0,0,0,0,0,0,0,0,1,-1,1,1), {\tt H}=(0,0,0,0,0,0,0,0,0,0,0,1,1),{\tt C}=(0,0,0,0,0,0,0,0,0,0,1,0,1),\break  {\tt D}=(0,0,0,0,0,0,0,0,0,0,1,0,-1), \ {\tt q}=(0,0,0,0,0,0,0,0,0,1,0,0,-1)

\subsection{\label{app:1k} 14-dim MMPHs}

{\bf 19-9}\quad {\tt 123456789ABCDE,12345679ABFGD,1OPF,27a,347E,3479ABP,567CG,9a,O89AB.}

\medskip

{\bf 66-15} (master for {\bf 19-9}) {\tt 123456789ABCDE,12345679ABFGDH,1IJKLMNOPFQRST,27UVWXMNYZTabc,347defghijkElm,\break 347defg9ABPHnm,347defgFijkGDH,567opqrOBPFsZt,567opqr9ABaunl,567opqrijbkCGu,opvwxyLX9iQYab,\break qdz!"KWyAt\#Sab,re\$JVx!"s\%abck,fgIUvwz\$Oh89AB, fgIUvwz\$\#R\%jab.} {\tt 1}=(0,0,1,1,1,1,0,0,0,0,0,0,0,0),\break  {\tt 2}=(0,0,1,-1,1,-1,0,0,0,0,0,0,0,0), {\tt 3}=(0,0,0,1,0,-1,0,0,0,0,0,0,0,0), {\tt 4}=(0,0,1,0,-1,0,0,0,0,0,0,0,0,0),\break  {\tt 5}=(0,1,0,0,0,0,0,0,0,0,0,0,0,0), {\tt 6}=(1,0,0,0,0,0,0,0,0,0,0,0,0,0), {\tt 7}=(0,0,0,0,0,0,1,0,0,0,0,0,0,0),\break  {\tt o}=(0,0,0,1,0,0,0,0,0,0,0,0,0,0), {\tt p}=(0,0,1,0,0,0,0,0,0,0,0,0,0,0), {\tt q}=(0,0,0,0,0,1,0,0,0,0,0,0,0,0),\break  {\tt r}=(0,0,0,0,1,0,0,0,0,0,0,0,0,0), {\tt d}=(1,-1,1,0,1,0,0,0,0,0,0,0,0,0), {\tt e}=(1,1,0,1,0,1,0,0,0,0,0,0,0,0),\break  {\tt f}=(1,1,0,-1,0,-1,0,0,0,0,0,0,0,0), {\tt g}=(-1,1,1,0,1,0,0,0,0,0,0,0,0,0), {\tt I}=(0,1,-1,1,0,0,1,0,0,0,0,0,0,0),\break  {\tt U}=(1,0,1,1,0,0,0,-1,0,0,0,0,0,0), {\tt v}=(1,0,0,0,1,1,0,1,0,0,0,0,0,0), {\tt w}=(0,1,0,0,-1,1,-1,0,0,0,0,0,0,0),\break  {\tt z}=(0,0,1,0,-1,0,1,1,0,0,0,0,0,0), {\tt \$}=(0,0,0,1,0,-1,-1,1,0,0,0,0,0,0), {\tt J}=(1,0,1,0,0,-1,1,0,0,0,0,0,0,0),\break  {\tt V}=(0,-1,1,0,0,1,0,1,0,0,0,0,0,0), {\tt x}=(-1,1,0,0,0,0,1,1,0,0,0,0,0,0), {\tt !}=(1,0,-1,-1,0,0,0,1,0,0,0,0,0,0),\break  {\tt "}=(0,1,1,-1,0,0,-1,0,0,0,0,0,0,0), {\tt K}=(1,0,0,1,-1,0,-1,0,0,0,0,0,0,0), {\tt W}=(0,1,0,1,1,0,0,1,0,0,0,0,0,0),\break  {\tt y}=(1,1,0,0,0,0,1,-1,0,0,0,0,0,0), {\tt L}=(0,1,0,0,1,-1,-1,0,0,0,0,0,0,0), {\tt X}=(1,0,0,0,-1,-1,0,1,0,0,0,0,0,0),\break  {\tt M}=(1,1,0,-1,0,1,0,0,0,0,0,0,0,0), {\tt N}=(1,-1,-1,0,1,0,0,0,0,0,0,0,0,0), {\tt O}=(0,0,0,0,0,0,0,0,0,0,1,0,0,0),\break  {\tt h}=(0,0,0,0,0,0,0,0,1,-1,0,0,0,0), {\tt 8}=(0,0,0,0,0,0,0,0,1,1,0,0,0,0), {\tt 9}=(0,0,0,0,0,0,0,0,0,0,0,0,0,1),\break  {\tt A}=(0,0,0,0,0,0,0,0,0,0,0,1,1,0), {\tt B}=(0,0,0,0,0,0,0,0,0,0,0,1,-1,0), {\tt P}=(0,0,0,0,0,0,0,1,0,-1,0,0,0,0),\break  {\tt F}=(0,0,0,0,0,0,0,1,0,1,0,0,0,0), {\tt i}=(0,0,0,0,0,0,0,0,0,0,0,1,0,0), {\tt Q}=(0,0,0,0,0,0,0,0,1,0,0,0,-1,0),\break  {\tt Y}=(0,0,0,0,0,0,0,0,1,0,0,0,1,0), {\tt s}=(0,0,0,0,0,0,0,0,1,0,0,1,1,-1), {\tt Z}=(0,0,0,0,0,0,0,0,1,0,0,-1,-1,-1),\break  {\tt t}=(0,0,0,0,0,0,0,0,1,0,0,0,0,1), {\tt \#}=(0,0,0,0,0,0,0,0,1,0,0,1,-1,-1), {\tt R}=(0,0,0,0,0,0,0,0,1,0,0,1,1,1),\break  {\tt \%}=(0,0,0,0,0,0,0,0,1,0,0,-1,0,0), {\tt j}=(0,0,0,0,0,0,0,0,0,0,0,0,1,-1), {\tt S}=(0,0,0,0,0,0,0,0,1,0,0,-1,1,-1),\break  {\tt T}=(0,0,0,0,0,0,0,0,0,0,0,1,0,-1), {\tt a}=(0,0,0,0,0,0,0,0,0,1,1,0,0,0), {\tt b}=(0,0,0,0,0,0,0,0,0,1,-1,0,0,0),\break  {\tt c}=(0,0,0,0,0,0,0,0,1,0,0,1,-1,1), {\tt k}=(0,0,0,0,0,0,0,0,0,0,0,0,1,1), {\tt C}=(0,0,0,0,0,0,0,1,-1,1,1,0,0,0),\break  {\tt G}=(0,0,0,0,0,0,0,1,-1,-1,-1,0,0,0), {\tt u}=(0,0,0,0,0,0,0,1,1,0,0,0,0,0), {\tt D}=(0,0,0,0,0,0,0,1,1,-1,1,0,0,0),\break  {\tt E}=(0,0,0,0,0,0,0,1,0,0,-1,0,0,0), {\tt H}=(0,0,0,0,0,0,0,0,1,0,-1,0,0,0), {\tt n}=(0,0,0,0,0,0,0,1,-1,1,-1,0,0,0),\break  {\tt l}=(0,0,0,0,0,0,0,1,-1,-1,1,0,0,0),\qquad\qquad {\tt m}=(0,0,0,0,0,0,0,1,1,1,1,0,0,0)

\subsection{\label{app:1l} 15-dim MMPHs}

{\bf 25-8} \ {\tt 123456789ABCDEF,1234567GHIJKLD,1RS89ABCDEF,27TRSX9GH,347CL,347BK,347AIJ,T8X9.}

\medskip 

{\bf 66-14} (master$\,\,$for$\,\,${\bf 25-8}) {\tt 123456789ABCDEF,1234567GHIJKLDM,1NOPQRS89ABCDEF,27TUVWRSX9YGHZa,\break 347bcdeZfghiCjL,347bcdeaklBmKhi,347bcdenoApIJgl,567qrstuYpmjMEF, 567qrstvw9Yfkno,qrxyz!QW8uvwX9Y,\break qrxyz!QWX9YGHZa,sb"\#\$PV!8uvwX9Y,tc\%OUz\#\$8uvwX9Y,deNTxy"\%8uvwX9Y.} {\tt 1}=(0,0,1,1,1,1.0,0,0,0,0,0,0,0,0),\break  {\tt 2}=(0,0,1,-1,1,-1.0,0,0,0,0,0,0,0,0), {\tt 3}=(0,0,0,1,0,-1.0,0,0,0,0,0,0,0,0), {\tt 4}=(0,0,1,0,-1,0.0,0,0,0,0,0,0,0,0),\break  {\tt 5}=(0,1,0,0,0,0.0,0,0,0,0,0,0,0,0), {\tt 6}=(1,0,0,0,0,0.0,0,0,0,0,0,0,0,0), {\tt 7}=(0,0,0,0,0,0.1,0,0,0,0,0,0,0,0),\break  {\tt q}=(0,0,0,1,0,0.0,0,0,0,0,0,0,0,0), {\tt r}=(0,0,1,0,0,0.0,0,0,0,0,0,0,0,0), {\tt s}=(0,0,0,0,0,1.0,0,0,0,0,0,0,0,0),\break  {\tt t}=(0,0,0,0,1,0.0,0,0,0,0,0,0,0,0), {\tt b}=(1,-1,1,0,1,0.0,0,0,0,0,0,0,0,0), {\tt c}=(1,1,0,1,0,1.0,0,0,0,0,0,0,0,0),\break  {\tt d}=(1,1,0,-1,0,-1.0,0,0,0,0,0,0,0,0), {\tt e}=(-1,1,1,0,1,0.0,0,0,0,0,0,0,0,0), {\tt N}=(0,1,-1,1,0,0.1,0,0,0,0,0,0,0,0),\break  {\tt T}=(1,0,1,1,0,0.0,-1,0,0,0,0,0,0,0), {\tt x}=(1,0,0,0,1,1.0,1,0,0,0,0,0,0,0), {\tt y}=(0,1,0,0,-1,1.-1,0,0,0,0,0,0,0,0),\break  {\tt "}=(0,0,1,0,-1,0.1,1,0,0,0,0,0,0,0), {\tt \%}=(0,0,0,1,0,-1.-1,1,0,0,0,0,0,0,0), {\tt O}=(1,0,1,0,0,-1.1,0,0,0,0,0,0,0,0),\break  {\tt U}=(0,-1,1,0,0,1.0,1,0,0,0,0,0,0,0), {\tt z}=(-1,1,0,0,0,0.1,1,0,0,0,0,0,0,0), {\tt \#}=(1,0,-1,-1,0,0.0,1,0,0,0,0,0,0,0),\break  {\tt \$}=(0,1,1,-1,0,0.-1,0,0,0,0,0,0,0,0), {\tt P}=(1,0,0,1,-1,0.-1,0,0,0,0,0,0,0,0), {\tt V}=(0,1,0,1,1,0.0,1,0,0,0,0,0,0,0),\break  {\tt !}=(1,1,0,0,0,0.1,-1,0,0,0,0,0,0,0), {\tt Q}=(0,1,0,0,1,-1.-1,0,0,0,0,0,0,0,0), {\tt W}=(1,0,0,0,-1,-1.0,1,0,0,0,0,0,0,0),\break  {\tt R}=(1,1,0,-1,0,1.0,0,0,0,0,0,0,0,0), {\tt S}=(1,-1,-1,0,1,0.0,0,0,0,0,0,0,0,0), {\tt 8}=(0,0,0,0,0,0.0,0,0,1,1,1,1,0,0),\break  {\tt u}=(0,0,0,0,0,0.0,0,0,1,-1,1,-1,0,0), {\tt v}=(0,0,0,0,0,0.0,0,0,0,1,0,-1,0,0), {\tt w}=(0,0,0,0,0,0.0,0,0,1,0,-1,0,0,0),\break  {\tt X}=(0,0,0,0,0,0.0,0,1,0,0,0,0,0,0), {\tt 9}=(0,0,0,0,0,0.0,0,0,0,0,0,0,0,1), {\tt Y}=(0,0,0,0,0,0.0,0,0,0,0,0,0,1,0),\break  {\tt G}=(0,0,0,0,0,0.0,0,0,0,1,0,0,0,0), {\tt H}=(0,0,0,0,0,0.0,0,0,1,0,0,0,0,0), {\tt Z}=(0,0,0,0,0,0.0,0,0,0,0,0,1,0,0),\break  {\tt a}=(0,0,0,0,0,0.0,0,0,0,0,1,0,0,0), {\tt f}=(0,0,0,0,0,0.0,1,-1,1,0,1,0,0,0), {\tt k}=(0,0,0,0,0,0.0,1,1,0,1,0,1,0,0),\break  {\tt n}=(0,0,0,0,0,0.0,1,1,0,-1,0,-1,0,0), {\tt o}=(0,0,0,0,0,0.0,1,-1,-1,0,-1,0,0,0), {\tt A}=(0,0,0,0,0,0.0,0,1,-1,1,0,0,1,0),\break  {\tt p}=(0,0,0,0,0,0.0,1,0,1,1,0,0,0,-1), {\tt I}=(0,0,0,0,0,0.0,1,0,0,0,1,1,0,1), {\tt J}=(0,0,0,0,0,0.0,0,1,0,0,-1,1,-1,0),\break  {\tt g}=(0,0,0,0,0,0.0,0,0,1,0,-1,0,1,1), {\tt l}=(0,0,0,0,0,0.0,0,0,0,1,0,-1,-1,1), {\tt B}=(0,0,0,0,0,0.0,1,0,1,0,0,-1,1,0),\break  {\tt m}=(0,0,0,0,0,0.0,0,1,-1,0,0,-1,0,-1), {\tt K}=(0,0,0,0,0,0.0,1,-1,0,0,0,0,-1,-1), {\tt h}=(0,0,0,0,0,0.0,1,0,-1,-1,0,0,0,1),\break  {\tt i}=(0,0,0,0,0,0.0,0,1,1,-1,0,0,-1,0), {\tt C}=(0,0,0,0,0,0.0,1,0,0,1,-1,0,-1,0), {\tt j}=(0,0,0,0,0,0.0,0,1,0,1,1,0,0,1),\break  {\tt L}=(0,0,0,0,0,0.0,1,1,0,0,0,0,1,-1), {\tt D}=(0,0,0,0,0,0.0,0,1,0,0,1,-1,-1,0), {\tt M}=(0,0,0,0,0,0.0,1,0,0,0,-1,-1,0,1),\break {\tt E}=(0,0,0,0,0,0.0,1,1,0,-1,0,1,0,0), \qquad\qquad {\tt F}=(0,0,0,0,0,0.0,1,-1,-1,0,1,0,0,0)

\subsection{\label{app:1m} 16-dim MMPHs}

{\bf 22-9} \quad {\tt 123456789ABCDEFG,17HID,28UdG,3478efE,5678,Bd,e9,fICF,HUA.}

\medskip

{\bf 70-9} (master for {\bf 22-9}) {\tt 123456789ABCDEFG,17HIJKLMNOPQRDST,28UVWXLMYZabScdG,3478efghijPbkElm,\break 5678nopqrZsQtukT,novwxyKXzasBu!dm,pe"\#\$JWy9iY\%tR\&',qf(IVx\#\$zOjC)lF\&,ghHUvw"(ArN\%)c!'}.\break  {\tt 1}=(0,0,1,1,1,1,0,0,0,0,0,0,0,0,0,0), {\tt 2}=(0,0,1,-1,1,-1,0,0,0,0,0,0,0,0,0,0), {\tt 3}=(0,0,0,1,0,-1,0,0,0,0,0,0,0,0,0,0),\break  {\tt 4}=(0,0,1,0,-1,0,0,0,0,0,0,0,0,0,0,0), {\tt 5}=(0,1,0,0,0,0,0,0,0,0,0,0,0,0,0,0), {\tt 6}=(1,0,0,0,0,0,0,0,0,0,0,0,0,0,0,0),\break  {\tt 7}=(0,0,0,0,0,0,0,1,0,0,0,0,0,0,0,0), {\tt 8}=(0,0,0,0,0,0,1,0,0,0,0,0,0,0,0,0), {\tt n}=(0,0,0,1,0,0,0,0,0,0,0,0,0,0,0,0),\break  {\tt o}=(0,0,1,0,0,0,0,0,0,0,0,0,0,0,0,0), {\tt p}=(0,0,0,0,0,1,0,0,0,0,0,0,0,0,0,0), {\tt q}=(0,0,0,0,1,0,0,0,0,0,0,0,0,0,0,0),\break  {\tt e}=(1,-1,1,0,1,0,0,0,0,0,0,0,0,0,0,0), {\tt f}=(1,1,0,1,0,1,0,0,0,0,0,0,0,0,0,0), {\tt g}=(1,1,0,-1,0,-1,0,0,0,0,0,0,0,0,0,0),\break  {\tt h}=(-1,1,1,0,1,0,0,0,0,0,0,0,0,0,0,0), {\tt H}=(0,1,-1,1,0,0,1,0,0,0,0,0,0,0,0,0), {\tt U}=(1,0,1,1,0,0,0,-1,0,0,0,0,0,0,0,0),\break  {\tt v}=(1,0,0,0,1,1,0,1,0,0,0,0,0,0,0,0), {\tt w}=(0,1,0,0,-1,1,-1,0,0,0,0,0,0,0,0,0), {\tt "}=(0,0,1,0,-1,0,1,1,0,0,0,0,0,0,0,0),\break  {\tt (}=(0,0,0,1,0,-1,-1,1,0,0,0,0,0,0,0,0), {\tt I}=(1,0,1,0,0,-1,1,0,0,0,0,0,0,0,0,0), {\tt V}=(0,-1,1,0,0,1,0,1,0,0,0,0,0,0,0,0),\break  {\tt x}=(-1,1,0,0,0,0,1,1,0,0,0,0,0,0,0,0), {\tt \#}=(1,0,-1,-1,0,0,0,1,0,0,0,0,0,0,0,0), {\tt \$}=(0,1,1,-1,0,0,-1,0,0,0,0,0,0,0,0,0),\break  {\tt J}=(1,0,0,1,-1,0,-1,0,0,0,0,0,0,0,0,0), {\tt W}=(0,1,0,1,1,0,0,1,0,0,0,0,0,0,0,0), {\tt y}=(1,1,0,0,0,0,1,-1,0,0,0,0,0,0,0,0),\break  {\tt K}=(0,1,0,0,1,-1,-1,0,0,0,0,0,0,0,0,0), {\tt X}=(1,0,0,0,-1,-1,0,1,0,0,0,0,0,0,0,0), {\tt L}=(1,1,0,-1,0,1,0,0,0,0,0,0,0,0,0,0),\break  {\tt M}=(1,-1,-1,0,1,0,0,0,0,0,0,0,0,0,0,0), {\tt 9}=(0,0,0,0,0,0,0,0,1,0,0,0,0,0,0,0),  {\tt A}=(0,0,0,0,0,0,0,0,0,1,0,0,0,0,0,0),\break {\tt i}=(0,0,0,0,0,0,0,0,0,1,1,0,0,0,0,0), {\tt Y}=(0,0,0,0,0,0,0,0,0,1,-1,0,0,0,0,0), {\tt r}=(0,0,0,0,0,0,0,0,1,0,1,0,0,0,0,0),\break  {\tt N}=(0,0,0,0,0,0,0,0,1,0,-1,0,0,0,0,0), {\tt z}=(0,0,0,0,0,0,0,0,1,-1,0,0,0,0,0,0), {\tt \%}=(0,0,0,0,0,0,0,0,0,0,0,1,0,0,0,0),\break  {\tt O}=(0,0,0,0,0,0,0,0,1,1,1,1,0,0,0,0), {\tt j}=(0,0,0,0,0,0,0,0,1,1,-1,-1,0,0,0,0), {\tt P}=(0,0,0,0,0,0,0,0,1,-1,1,-1,0,0,0,0),\break  {\tt Z}=(0,0,0,0,0,0,0,0,1,-1,-1,-1,0,0,0,0), {\tt a}=(0,0,0,0,0,0,0,0,1,1,1,-1,0,0,0,0),  {\tt s}=(0,0,0,0,0,0,0,0,1,1,-1,1,0,0,0,0),\break  {\tt b}=(0,0,0,0,0,0,0,0,1,0,0,1,0,0,0,0), {\tt B}=(0,0,0,0,0,0,0,0,0,0,1,1,0,0,0,0),  {\tt C}=(0,0,0,0,0,0,0,0,0,0,1,-1,0,0,0,0),\break  {\tt Q}=(0,0,0,0,0,0,0,0,0,1,0,-1,0,0,0,0), {\tt t}=(0,0,0,0,0,0,0,0,0,0,0,0,1,0,0,0),  {\tt R}=(0,0,0,0,0,0,0,0,0,0,0,0,0,1,0,0),\break  {\tt u}=(0,0,0,0,0,0,0,0,0,0,0,0,0,1,1,0), {\tt k}=(0,0,0,0,0,0,0,0,0,0,0,0,0,1,-1,0),  {\tt D}=(0,0,0,0,0,0,0,0,0,0,0,0,1,0,1,0),\break  {\tt S}=(0,0,0,0,0,0,0,0,0,0,0,0,1,0,-1,0), {\tt )}=(0,0,0,0,0,0,0,0,0,0,0,0,1,-1,0,0),  {\tt T}=(0,0,0,0,0,0,0,0,0,0,0,0,0,0,0,1),\break  {\tt c}=(0,0,0,0,0,0,0,0,0,0,0,0,1,1,1,1), {\tt !}=(0,0,0,0,0,0,0,0,0,0,0,0,1,1,-1,-1),  {\tt d}=(0,0,0,0,0,0,0,0,0,0,0,0,1,-1,1,-1),\break  {\tt E}=(0,0,0,0,0,0,0,0,0,0,0,0,1,-1,-1,-1), {\tt l}=(0,0,0,0,0,0,0,0,0,0,0,0,1,1,1,-1),  {\tt F}=(0,0,0,0,0,0,0,0,0,0,0,0,1,1,-1,1),\break  {\tt m}=(0,0,0,0,0,0,0,0,0,0,0,0,1,0,0,1), {\tt \&}=(0,0,0,0,0,0,0,0,0,0,0,0,0,0,1,1),  {\tt '}=(0,0,0,0,0,0,0,0,0,0,0,0,0,0,1,-1),\break {\tt G}=(0,0,0,0,0,0,0,0,0,0,0,0,0,1,0,-1)

\end{document}